\documentclass[aps,prd,twocolumn,groupedaddress,showpacs,superscriptaddress]{revtex4-1}
\usepackage{aas_macros}

\usepackage{sidecap}

\usepackage{textcomp}
\usepackage{gensymb}
\usepackage{graphicx}
\usepackage{dcolumn}
\usepackage{bm}
\usepackage{amsmath}
\usepackage{afterpage}
\usepackage{tikz}
\usepackage{tikz-3dplot}
\usepackage[colorlinks,citecolor=blue]{hyperref}

\usepackage{color}

\begin{document}

\preprint{APS/123-QED}

\title{Self-lensing flares from black hole binaries I:\\ general-relativistic ray tracing of black hole binaries}

\author{Jordy Davelaar}
 \email{jrd2210@columbia.edu}
\affiliation{Department of Astronomy, Columbia University, 550 W 120th St, New York, NY 10027, USA}
\affiliation{Columbia Astrophysics Laboratory, Columbia University, 550 W 120th St, New York, NY 10027, USA}%
\affiliation{Center for Computational Astrophysics, Flatiron Institute, 162 Fifth Avenue, New York, NY 10010, USA}%

\author{Zolt\'an Haiman}
\affiliation{Department of Astronomy, Columbia University, 550 W 120th St, New York, NY 10027, USA}

\date{\today}

\begin{abstract}
The self-lensing of a massive black hole binary (MBHB), which occurs when the two BHs are aligned close to the line of sight, is expected to produce periodic, short-duration flares. Here we study the shapes of self-lensing flares (SLFs) via general-relativistic ray tracing in a superimposed binary BH metric, in which the emission is generated by geometrically thin accretion flows around each component. The suite of models covers eccentric binary orbits, black hole spins, unequal mass binaries, and different emission model geometries. We explore the above parameter space, and report how the light curves change as a function of, e.g., binary separation, inclination, and eccentricity. We also compare our light curves to those in the microlensing approximation, and show how strong deflections, as well as time-delay effects, change the size and shape of the SLF. If gravitational waves (GWs) from the inspiraling MBHB are observed by LISA, SLFs can help securely identify the source and localizing it on the sky, and to constrain the graviton mass by comparing the phasing of the SLFs and the GWs. Additionally, when these systems are viewed edge-on the SLF shows a distinct dip that can be directly correlated with the BH shadow size.  This opens a new way to measure BH shadow sizes in systems that are unresolvable by current VLBI facilities.
\end{abstract}


\maketitle

\section{\label{sec:intro} Introduction}

Massive black hole binaries (MBHB) are thought to reside in the nuclei of numerous galaxies as a result of galaxy mergers \citep{begelman1980}. As their orbits shrink, they will eventually merge due to gravitational wave radiation. MBHBs are also the primary candidates to be observed by LISA~\cite{LISA}, and are targeted by searches for gravitational waves (GWs) via pulsar timing arrays (PTAs; \cite{NANOGrav12.5}).  MBHBs are expected to be surrounded by a gaseous circumbinary disk, from which material is accreted towards both black holes (BHs), forming a so-called mini-disk around each component. In many cases, the accretion rates are expected to be close to the Eddington limit, and the mini-disks are then best described by geometrically thin and optically thick accretion flows. Electromagnetic emission from these systems should be detectable starting well before the merger, and should persist all the way to the merger~\citep{farris+2015b,tang2018}.

Observational evidence for compact MBHBs was, however, until recently sparse (see, e.g.~\cite{derosa2019,bogdanovic+2021} for comprehensive recent reviews). The first detected candidates are large-separation binaries of several kpc \citep{dotti2012,comerford2013}. More recently, with large optical time-domain surveys, several active galactic nuclei (AGN) have been identified that show quasi-periodic behavior in their light curves \citep{graham2015,charisi2016,liu2019,liu+2020,chen+2020}.

This periodicity can be attributed either to periodic hydrodynamical modulations of the accretion flow \citep{artymowicz1994,macfadyen2008,dorazio2013,shi2015,bowen2019}, or to relativistic Doppler effects \citep{dorazio2015,hu2020}. One Kepler source KIC-11606854 \citep{hu2020}, also known as Spikey, shows, in addition, a short duration flare, lasting for a time scale of an hour. This short duration makes an accretion-induced flare scenario unlikely, since the viscous and the orbital timescales in the disk regions believed to be responsible for optical emission are both too long compared to the flaring time. An alternative mechanism for these flares is self lensing of the MBHB \citep{dorazio2018,hu2020,ingram2021}.

Assuming the BHs orbit on elliptical Keplerian orbits, self-lensing can occur when the binary is viewed close to edge-on. For a distant observer, one of the mini-disks is lensed by the other BH twice per orbit. Lensing occurs when the two BHs are aligned with respect to the line of sight, such that the source is within the Einstein angle of the lens. The Einstein angle for a point-mass lens is given by~\citep[e.g.][]{gaudi2012}

\begin{equation}
    \theta_{\rm E} = r_{\rm E}/D =\sqrt{\frac{4GM}{Dc^2}},
\end{equation}
where $r_{\rm E}$ is the Einstein radius, $G$ is Newton's constant, $M$ is the mass of the lens, $D^{-1}\equiv D_L^{-1}-D_S^{-1}$, $D_L$ and $D_S$ are the distances to the lens and the source, respectively, and $c$ is the speed of light.

Previous work has raised the possibility of such self-lensing, using simple models~\citep{jaroszynski1992,beky2013,haiman2017,dorazio2018,hu2020}. In these studies, lensing was approximated in the limit of microlensing, meaning that the lens and the source are both taken to be point-like, and the deflections angles are assumed to be small~\cite{paczynski1986}. These simple models demonstrated that self-lensing, together with the Doppler modulation from the orbital motion, can produce the observed sinusoidal trends in the light curves, as well as produce recurring lensing flares for nearly edge-on binaries. However, they did not consider the effects of finite light travel time and photon trajectories in strongly relativistic spacetimes. Additionally, \cite{haiman2017} and \cite{hu2020} both assume the emission from each component to be point-like.

Ref.~\cite{dorazio2018} presented, in addition, the first exhaustive study, including the dependence of the flare shapes on source morphology by including a finite-sized disk emission model. The mini-disks were assumed to extend from the the innermost stable circular orbit (ISCO) to the tidal truncation radius. On the other hand, these models still utilized the microlensing approximation, for each disk patch.

The strong bending of photon trajectories in the strongly relativistic regions can introduce new effects in two ways. First, in the "source", the lensing of the mini-disk emission by its own central BH warps the emission region around it, enlarging its apparent size on the sky and strongly distorting its shape.  Second, for sufficiently compact binaries, rays passing near the other BH (i.e. the "lens") can suffer further large deflections.

To our knowledge, the first mentions of MBHB self-lensing modeled with general relativistic ray tracing (GRRT) was made by \citep{pihajoki2018}, who computed the emission from thin Novikov-Thorne disks in curved spacetimes. They modeled a binary with a mass ratio of $q=0.01$ and showed a single light curve. See also related works by \citep{bohn2015} who ray traced numerical binary metrics with a far away artificial screen, on ray tracing GRMHD MBHB simulations by ~\citep{kelly2017,dascoli2018}, on emission from binary neutron stars~\cite{schnittman2018}, and \citep{beky2013} on lensing by stellar transits around a supermassive BH. More recently, \citep{ingram2021} studied lensing flares by ray-tracing the image of the disk around the background BH  that is lensed by the other (foreground) BH. This method allowed them to compute light curves that include strong lensing distortions of the source morphology. This work, however, does not use a binary metric, only considers circular equal mass binaries, and does not include Doppler modulations due to orbital motion.

In this work, we expand on these recent studies by including the above, previously neglected, effects, and by exploring how the light curves of self-lensing binaries depend on numerous model parameters. In \S~\ref{sec:meth}, we construct an approximate superposed Cartesian binary metric, explain our adaptive general-relativistic ray-tracing code, and describe our semi-analytical emission models. In \S~\ref{sec:res}, we report the results of our parameter exploration, which are further discussed in \S~\ref{sec:dis}.  Finally in \S~\ref{sec:concl}, we summarize our main findings and the implications of this work.

\section{\label{sec:meth}Methods}

In this section, we introduce our approximate superposed binary metric and the Kepler orbits used for the binary, explain our adaptive general-relativistic ray-tracing code, and finally describe our emission models and introduce our model parameters.

\subsection{Metric}
For constructing an approximate binary metric, we use a superposition of two Cartesian Kerr-Schild metrics \citep{kerr1963}, this approach is identical to Ref. \citep{pihajoki2018}. Our covariant superposed metric, $g_{\mu\nu}$, in geometric units $G=M=c=1$, is given by,
\begin{equation}
g_{\mu\nu} = \eta_{\mu\nu} + h^p_{\mu\nu} + h^s_{\mu\nu},
\end{equation}
where $(p/s)$ superscripts indicate the primary or secondary BH, and $q$ is the mass ratio $q\equiv M_{\rm s}/M_{\rm p}\leq 1$. The metric is a superposition of the Minkowski metric $\eta_{\mu\nu}$ defined as ${\eta_{\mu\nu}={(-1,1,1,1)}}$, and two source terms $h^{p/s}_{\mu\nu}=f^{p/s} l_\mu^{p/s} l_\nu^{p/s}$ for the BHs, where $f$ is a scaling factor, and $l_\nu$ is a Killing vector. The source terms are shifted with respect to the original spatial coordinates $\vec{x}$ via $\vec{x}_{p/s} =  \vec{x} - \vec{x}^{bh}$ with $\vec{x}^{bh}_{p/s}$ the position vector of the BH.
The factor $f$ and Killing vector $l_\nu$ are given by,

\begin{eqnarray}
f &= \frac{2r^3}{r^4 + a^2 z^2},\\
l_\nu &=  \begin{pmatrix}
1\\
\frac{rx+a y}{r^2 + a^2} \\ \frac{ry-ax}{r^2 + a^2} \\ \frac{z}{r},
\end{pmatrix}
\end{eqnarray}
where $r$ is the radial coordinate, equivalent to the radius in spherical Kerr-Schid coordinates, given by
\begin{eqnarray}
r^2_ =& {\frac{R^2 - a^2 + \sqrt{(R^2 - a^2)^2 + 4a^2z^2}}{2} },\\
R^2 =& x^2+y^2+z^2.
\end{eqnarray}
All terms $f$ and $l^\nu$ are either using the variables for the primary or secondary BH. The contravariant metric is similarly defined as
\begin{equation}
g^{\mu\nu} = \eta^{\mu\nu} - h^{\mu\nu,p} - h^{\mu\nu,s},
\end{equation}
with $h^{\mu\nu,p/s}=f^{p/s} l^{\mu,p/s} l^{\nu,p/s}$. Here $l^{\nu,p/s}$ is identical to $l_\nu^{p/s}$ except that the temporal component changes sign.

The choice for a superposed metric violates the non-linearity of the Einstein equation. However, \citep{combi2021} showed that a superposed metric in post-Newtonian (PN) harmonic coordinates recovers the behavior of more extensive PN approximate metrics. In this work, we only consider relatively wide binaries, i.e. with separation $d$ much larger than $r_{\rm g}$ and $v \gamma < 1$, where $R_{\rm g}=\frac{GM}{c^2}$ is the gravitational radius, $v$ the orbital velocity, and $\gamma=1/\sqrt{1-v^2}$ the Lorentz factor.

In this limit \citep{combi2021} showed that the addition of "fake" mass, not accounted for in the source terms of the Einstein equations, is small. The benefit of a superposed metric is that it is computationally cheap, and therefore ideal for performing our parameter study as long the binaries have large separations, where, e.g., tidal deformations and gravitational radiation remain small corrections. In our case, we will limit ourselves to large separations of at least $100 R_g$ which also ensures that the orbital velocity remains at most mildly relativistic.

\subsection{Keplerian orbits}

Our assumed superposed metrics take the BH positions as an input. We assume the BHs to be on Keplerian orbits with eccentricity $e$. To find the position of the BHs given a separation $a_{per}$ at periapsis, total binary mass $M$, mass ratio $q$, eccentricity $e$, and time $t$, we  solve the Kepler equations in the center-of-mass frame following~\cite{murray2010}. We initially orient coordinate axes such that the orbital angular momentum vector of the binary is aligned along the $z$-axis,

which is further assumed parallel to the individual BH spin axes (although we will relax the latter assumption below by rotations the BH spin axis and the orbital axis). In more detail, to find the positions of the BHs as a function of time, we use the following steps:
\begin{enumerate}
 \item Given the periapsis distance, find the semi-major axis $a_{\rm major} = a_{\rm per}/(1-e)$ and compute the orbital period $T= \sqrt{4 \pi^2  a_{\rm major}^3 / (1 + q)}$.
 \item Given the period $T$ and the input time $t$, define the phase $n = 2 \pi t/T$.
 \item Solve Kepler's equation, $n(t-t_0)=E-e\cos E$ for the eccentric anomaly $E$, using a Newton-Raphson algorithm (here $t_0$ sets the pericenter passage time).
 \item Given $E$, find the radial distance {$r = a_{\rm major} (1-e \cos E)$}
 \item Given $r$, find the phase angle $f=t \sqrt{1+q}/r^{3/2}$ (for $e=0$) or $f=\cos^{-1}[(a_{\rm major} (1 - e^2) / r - 1)/e]$ (for $e\neq0$).
\end{enumerate}

\begin{figure}

\begin{tikzpicture}[scale=5,tdplot_main_coords]
    \coordinate (O) at (0,0,0);
    \draw[thick,->] (0,0,0) -- (1,0,0) node[anchor=north east]{$y$};
    \draw[thick,->] (0,0,0) -- (0,0.75,0) node[anchor=north west]{$x$};
    \draw[thick,->] (0,0,0) -- (0,0,0.75) node[anchor=south]{$z$};

    \begin{scope}[canvas is yx plane at z=0]
        \filldraw[opacity=0.05,color=gray] (0,0) circle (0.5cm);
        \draw[color=gray] (0,0) circle (0.5cm);
    \end{scope}

    \definecolor{col1}{RGB}{37,52,148}
    \definecolor{col2}{RGB}{65,182,196}

    \tdplotsetrotatedcoords{315}{20}{0}
    \begin{scope}[tdplot_rotated_coords,canvas is xy plane at z=0]
        \filldraw[opacity=0.05,color=blue] (0,0) circle (0.5cm);
        \draw (0,0) circle (0.5cm);
    \end{scope}
 \draw[tdplot_rotated_coords,thick,->] (0,0,0) -- (0,0,0.9)node[anchor=south]{$\vec{L}_{\rm orbit}$} ;

    \draw[thick,color=col2] (-0.35,-0.35,0) -- (0.35,0.35,0)node[anchor=west,rotate=-45,outer sep=-5pt,xshift=-80pt,yshift=-5pt]{Nodes} ;
    \tdplotdrawarc[dashed,color=col2]{(O)}{0.20}{0}{45}{anchor=north}{$\Omega$}

    \draw[thick,color=col1] (0,0,0) -- (-0.32,0.32,0.2) ;
    \draw[dashed, color=col1] (-0.32,0.32,0) -- (-0.32,0.32,0.2);
    \draw[dashed, color=col1] (0,0,0) -- (-0.32,0.32,0);

    \tdplotsetthetaplanecoords{130}
    \tdplotdrawarc[tdplot_rotated_coords,color=col1]{(0,0,0)}{0.25}{62}{88}{anchor=south west}{$I$}

    \begin{scope}[canvas is xy plane at z=0.15]
        \filldraw[opacity=0.5,color=orange] (0,0.49) circle (0.1cm);
        \draw (0,0.49) circle (0.1cm);
    \end{scope}

    \begin{scope}[canvas is xy plane at z=-0.15]
        \filldraw[opacity=0.5,color=orange] (0,-0.49) circle (0.1cm);
        \draw (0,-0.49) circle (0.1cm);
    \end{scope}

    \filldraw[color=black] (0,0.49,0.15) circle (0.025cm);

    \filldraw[color=black] (0,-0.49,-0.15) circle (0.025cm);

    \tdplotsetcoord{P}{0.75}{80}{90}
    \draw[color=gray,->] (O) -- (P) node[above right] {observer};
    \tdplotsetcoord{P}{0.74}{80}{90}
    \draw[dashed, color=gray] (P) -- (Pxy);

    \tdplotsetthetaplanecoords{90}
     \tdplotdrawarc[tdplot_rotated_coords,color=gray]{(0,0,0)}{0.7}{-23}%
         {80}{anchor=south west}{$i_{\rm orbit}$}
\end{tikzpicture}
\caption{Illustration of the model setup. The binary's center of mass is at the origin, and the observer is located in the $x-z$ plane. The binary's orbital plane is tilted with respect to the observer's line of sight by the angle $i_{\rm orbit}$.  The BH spin axes and the orbital angular momentum vectors of the minidisks are kept parallel to the $z$-axis, and are misaligned with the binary's orbital plane by the angle $I$.  The node angle $\Omega$ specifies the orientation of the semi-major axis of the elliptical binary orbit within the orbital plane.}
    \label{fig:cartoon}
\end{figure}
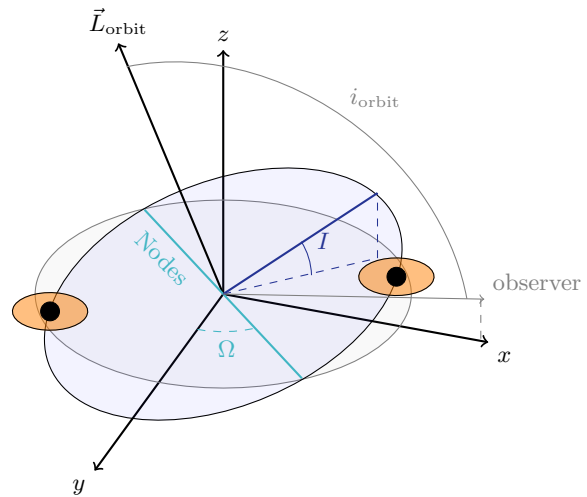

The position vector is then given by
\begin{equation}
  \vec{X}=
\begin{pmatrix}
X \\
Y \\
Z
\end{pmatrix}
=
\begin{pmatrix}
r \cos f \\
r \sin f \\
0
\end{pmatrix}.
\end{equation}
Finally, we introduce a rotation of the binary around the $z$-axis by a node angle $\Omega$ and around the y axis by the inclination angle $I$. The rotation around the y axis keeps the BH spin axes fixed in the z direction. The angular momentum vectors of both minidisks are also kept aligned with BH spin axes, and are in this case misaligned with the binary orbit. The inclination angle of the observer is defined with respect to the angular momentum vector of the binary, with the observer located in the $x-z$ plane, as illustrated in Fig.~\ref{fig:cartoon}. This results in the following rotated position vector,
\begin{equation}
  \vec{X}=
\begin{pmatrix}
r \cos(f+\Omega)\cos I \\
r \sin(f+\Omega) \\
r \cos(f+\Omega)\sin I
\end{pmatrix}.
\end{equation}
The positions of the individual BHs are then given by ${\vec{x}^{p} = \frac{1}{1+q} \vec{X}}$ and ${\vec{x}^{s} = -\frac{q}{1+q} \vec{X}}$.

The velocity of the binary is computed by taking the time derivate of the position vector and is given by
\begin{equation}
  \vec{V}=
\begin{pmatrix}
(\dot{r} \cos(f+\Omega) - r \dot{f}\sin(f+\Omega) )\cos I \\
\dot{r} \sin(f+\Omega) + r \dot{f}\cos(f+\Omega) ) \\
(\dot{r} \cos(f+\Omega) - r \dot{f}\sin(f+\Omega) )\sin I
\end{pmatrix}
\end{equation}
where
\begin{eqnarray}
    \dot{r} = 2\pi \frac{a_{\rm major} e  \sin f}{ T \sqrt{1 - e^2}} \\
    r\dot{f} = 2\pi \frac{a_{\rm major}(1 + e  \cos f)}{ T \sqrt{1 - e^2}}
\end{eqnarray}
The BH velocity vector is then given by $\vec{v}^{p} = \frac{1}{1+q} \vec{V}$ and $\vec{v}^{s} = - \frac{q}{1+q} \vec{V}$. These velocities are used to compute the Doppler shift of the observed frequency via, $\nu_{\rm obs} = \nu_{\rm emitted} /(\gamma (1-v_\parallel))$, where $\gamma=1/\sqrt{1-v^2}$ is the Lorentz factor,
and $v_\parallel$ the velocity parallel to the line of sight.

\tdplotsetmaincoords{60}{110}
\pgfmathsetmacro{\rvec}{.7}
\pgfmathsetmacro{\thetavec}{40}
\pgfmathsetmacro{\phivec}{90}
\def\roll{0}
\def\pitch{0}
\def\yaw{0}

\subsection{Adaptive general-relativistic ray tracing}
To generate synthetic images and light curves, we adapted the general-relativistic ray tracing (GRRT) code {\tt RAPTOR} \citep{bronzwaer2018,bronzwaer2020}. The code integrates the geodesic equation and simultaneously solves the radiation transport equation for multiple frequencies. The geodesic equation is given by,
\begin{equation}
 \frac{{\rm d}^2 x^{\alpha}}{{\rm d}\lambda^2} = -\Gamma^{\alpha}_{\ \mu \nu} \frac{{\rm d}x^{\mu}}{{\rm d}\lambda} \frac{{\rm d}x^{\nu}}{{\rm d}\lambda},
\end{equation}
with $\lambda$ is the affine parameter and $\Gamma^{\alpha}_{\ \mu\nu}$ the connection coefficients given by,
\begin{equation}
 \Gamma^{\alpha}_{\ \mu \nu} = \frac{1}{2} g^{\alpha \rho} \left[
 \partial_{\mu} g_{\nu \rho} + \partial_{\nu} g_{\mu \rho} -
 \partial_{\rho} g_{\mu \nu} \right].
 \label{eqn:christoffels}
\end{equation}
The connection coefficients depend on the derivates of the metric. These derivates are numerically computed by the code using a finite difference method.

As the code integrates the geodesic equation, it simultaneously solves the radiative transfer equation backward in time, which is given by

\begin{equation}
    \frac{d I_\nu}{\nu^3 d\lambda} = j_\nu \exp(-\tau_\nu).
\end{equation}
Here $j_\nu$ is the local specific emissivity coefficient and $\tau_\nu$ is the optical depth, defined as $\tau_\nu = \int_0^{\lambda_{\rm current}} \nu \alpha_\nu d\lambda$. The integral goes from the "camera" to the current point of integration, and $\alpha_\nu$ is the specific absorption coefficient which depends on the emission model (described in the following section). During the integration of a geodesic, we keep track of the coordinate time and change the positions of the BHs correspondingly. This introduces a retarded time in our models, often referred to as ``slow-light'' in the ray-tracing literature. This is in contrast to ``fast light'', where all dynamical processes are ignored, i.e. the metric as well as the plasma is assumed static during the ray-tracing (effectively corresponding to an infinite speed of light).

Light rays are initialized as geodesics attached to a pixel on a virtual camera held by the observer. We use a covariant tetrad camera described in \citep{davelaar2018}, positioned far away from the center of mass, $d_{\rm cam}=10^4 R_{\rm g}$. Since the binaries studied in this work have large separations and high resolution is needed only close to the BHs or to the Einstien ring, using a purely uniform resolution camera is computationally inefficient since most of the field of view is empty. To this end, we speed up our code by implementing a quadtree adaptive mesh refinement scheme on the camera plane, enabling the code to add resolution during run time only in the regions of interest. Our method differs from the approach by \citep{gelles2021}, who introduced adaptive gridding by using recursive subdivision of the image plane which is not block based. Our method is closer to adaptive mesh refinement strategies as used in GRMHD, e.g. \cite{porth2017}. The code initializes a uniform camera grid at relatively low resolution, consisting of $N_1 \times M_1$ blocks with $n \times m$ pixels. Given a predefined field of view (fov) of the total image fov$_x$ by fov$_y$, each block has a fov of block\_fov$_x$=  fov$_x$ /$N_1$ and block\_fov$_y$=fov$_y$/$M_1$, with pixel separation of dx=block\_fov$_x$/$n$ and dy=block\_fov$_y$/$m$. Each block has a unique set of indices $i,j$ from which we can compute the coordinates of the lower-left corner
\begin{eqnarray}
    {\rm lcorner}_x (i) =& - {\rm fov}_x /2 + i  {\rm block\_fov}_x\\
    {\rm lcorner}_y (j) =& - {\rm fov}_y /2 + j  {\rm block\_fov}_y,
\end{eqnarray}
and assign an initial set of impact parameters $\alpha,\beta$ for the pixel given a set of indices $k,l$,
\begin{eqnarray}
    \alpha =& {\rm lcorner}_x (i) + (k+0.5) dx\\
    \beta =& {\rm lcorner}_y (j) + (l+0.5) dy. \label{impact}
\end{eqnarray}
Every block is then ray-traced by the code, and the total intensity of each pixel is computed. The code then computes the sum $s_{ij}$ of the relative differences for every pixel with index $(k,l)$ in a block and their four direct neighbors (excluding diagonals) via $s_{kl}=\sum_{mn} \mid I_{kl}-I_{mn}\mid / I_{kl}$, and when the maximum of this sum in a block exceeds a user-defined threshold $s_{\rm th}$, refinement is triggered for the whole block. For $s_{\rm th}$ we use a value of $0.2$, corresponding with an increase of more than 20 percent between adjacent pixels. When refinement is triggered, the block splits in two in both directions, the level is incremented by one, and new indices for the new blocks are computed via
\begin{eqnarray}
    i =& &2 i_{\rm parent} + k_{\rm child}\mod{2}\\
    j =& &2 j_{\rm parent} + k_{\rm child} / 2,
\end{eqnarray}
where $k_{\rm child}$ is an index that runs from zero to four and $i_{\rm parent}$ and $j_{\rm parent}$ are the original $(i,j)$ indices of the parent block.
The coordinates of the lower-left corners of the new blocks are computed via
\begin{eqnarray}
    {\rm lcorner}_x (i) =& - {\rm fov}_x /2. + i   {\rm block\_fov}_x/2^{{\rm level}-1}\\
    {\rm lcorner}_y (j) =& - {\rm fov}_y /2. + j   {\rm block\_fov}_y/2^{{\rm level}-1},
\end{eqnarray}
where ${\rm level}$ is the current resolution level, from which the impact parameters can be computed from Eq.~\ref{impact}.

For each pixel in the block, the code then recomputes the emission at the new resolution level. This procedure repeats every time the refinement criterion is met or until the level of refinement of a block reaches a user-defined maximum.

\begin{figure}[ht!]
  \centering
  \includegraphics[width=0.48\textwidth]{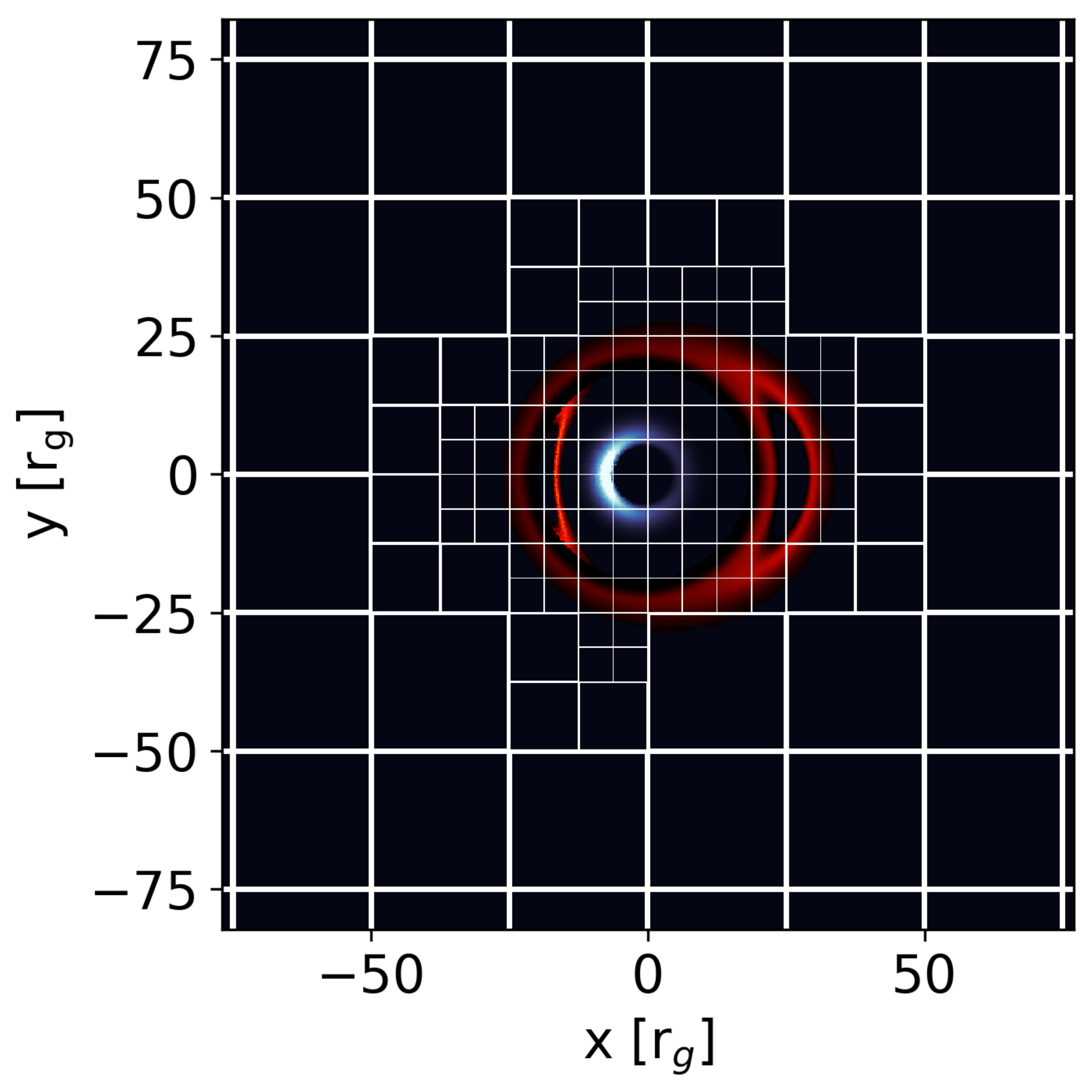}
    \caption{ Example of a ray-traced image in one of our models, illustrating the adaptive grid of the camera pixelization. White lines indicate blocks of equal resolution, each containing $25^2$ pixels. Higher resolution blocks are triggered only when a block overlaps with the source.}
    \label{fig:AMR}
\end{figure}

A high-resolution example of an adaptive image is shown in Fig. \ref{fig:AMR}, where white lines indicate blocks. The speedup obtained by this method depends on the image morphology. If an Einstein ring is present, the emission region is enlarged, and more high resolution blocks are triggered. In this case, the speedup is approximately a factor three, while in the absence of an Einstein ring, the speedup is approximately a factor ten.

Additionally, we can predict the location of the BHs on the image by solving Kepler's equation for the expected arrival time of the light ray at the BH. The code then only computes the blocks that are close to the predicted positions. This results in an additional factor of four speedup for our widest binaries ($a_{\rm major} = 1000 R_g$).
\subsection{Emission model}
To describe the emission from the plasma around each BH component, we adopt multicolor "minidisks", generating radiation by an optically thick and geometrically thin accretion flow. The disk resides in the $x$,$y$ plane as illustrated in Fig.~\ref{fig:cartoon},
meaning that the disk's angular momentum vector is parallel to the BH's spin axis. The disk extends from an inner radius $R_{\rm inner}$ up to the tidal radius. The inner radius is either at the event horizon, $R_{\rm h} = 1 + \sqrt{1-a^2}$ or at the ISCO, given by

\begin{eqnarray}
    R_{\rm ISCO} =  3 + Z_2 + \sqrt{(3-Z_1)(3+Z_1+2Z_2)}, \\
    Z_1 = 1 + \sqrt[3]{1-a^2} \left(\sqrt[3]{1+a} + \sqrt[3]{1-a}\right),\\
    Z_2 = \sqrt{3a^2+Z_1^2},
\end{eqnarray}
where $a$ is the dimensionless spin parameter of the BH. The tidal truncation radius of a minidisk is given approximately by \citep{roedig2014}
\begin{eqnarray}
    R_{\rm tidal, p} =& 0.27 q^{-0.3} a_{\rm major} (1-e),\\
    R_{\rm tidal, s} =& 0.27 q^{0.3} a_{\rm major} (1-e).
\end{eqnarray}

When a geodesic crosses the disk, the intensity is computed from a black body spectrum. The temperature of the black body spectrum is set via $T=(F/\sigma_{\rm b})^{1/4}$ where $\sigma_{\rm b}$ is the Boltzmann constant and $F$ the isotropic bolometric intensity (flux per unit surface area) at a given radius given by $F = 3 G M \dot{M}/ 8 \pi r^3$.

The velocity profile is assumed either to be circular, or additionally includes a radial free-fall velocity component. The latter is used in the variants of our model, discussed below, in which we set the inner radius to be smaller than the ISCO. Purely circular Kepler orbits inside the ISCO are non-physical and hence the accretion flow in the inner regions will have significant radial velocities. The spatial component of the four-velocity is given, in Boyer-Linquist coordinates, by
\begin{eqnarray}
    u^r &=&
    \begin{cases}
      0, & \text{if}\ R_{\rm inner}=R_{ISCO} \\
      -\sqrt{2 r (a^2 + r^2)} \Delta / \Sigma; &  \text{otherwise}
    \end{cases}\\
    u^\theta &=& 0.\\
    u^\phi &=& u^t \omega
\end{eqnarray}
where $\Sigma=(r^2 + a^2)^2 - a^2 \Delta \sin(\theta)^2$, $\Delta=r^2 + a^2 - 2 r$, and $\omega = 1/(r^{3/2} + a^{1/2})$ is the angular velocity, where $a$ is the BH spin. The time component of the four-velocity is then computed in such a way that it ensures the right normalization $u^\mu u_\mu = -1$ of the four-velocity, via
\begin{eqnarray}
     u^t =& \sqrt{(-1 - u^r u^r g_{rr}) / \Psi},\\
    \Psi =& g_{tt} + 2  \omega_{\rm tot}  g_{t\phi} + \omega_{\rm tot}^2  g_{\phi \phi}.
\end{eqnarray}
here the metric components are in Boyer-Lindquist coordinates and are given by
\begin{eqnarray}
    g_{tt} =& -\left(1 - \frac{2 r }{\rho}\right) \\
    g_{rr} =& \Sigma/\Delta \\
    g_{t\phi} =&  - \frac{2 a r \sin^2\theta}{\rho}\\
    g_{\phi\phi} =& \left(r^2 + a^2 + \frac{2 a^2  r  \sin^2\theta}{ \rho}\right)  \sin^2\theta\\
    \rho =& r^2 + a ^2  \cos^2\theta
\end{eqnarray}
The four-velocity in BL coordinates is then transformed to Cartesian Kerr-Schild coordinates.

\begin{table}[ht!]
\begin{tabular}{l|l|llll|llll|l}
 & Disk & Orbit &  &  &  & Black & hole  & &  & Opacity  \\
 \hline
 & $R_{\rm inner}$ &  $a_{\rm per}$ & $e$ & $i_{\rm orbit}$  & $\Omega$ & $M_p$ & $q$ & $a$ & $i_{\rm BH}$ &  $\tau$    \\
 \hline
  \multicolumn{11}{c}{Fiducial model}\\
  \hline
M0 & $r_{h}$ & 100 & 0  & 90 & 0 & $10^7$ & 1 & 0 & 90 & thick    \\
\hline
  \multicolumn{11}{c}{Inner radius dependence}\\
\hline

M1 & \bm{$r_{ISCO}$} & 100 & 0  & 90 & 0 & $10^7$ & 1 & 0 & 90 & thick    \\
\hline
  \multicolumn{11}{c}{Binary orbital inclination dependence}\\
  \hline

M2a & $r_{h}$ & 100 & 0  & \bf{89} & 0 & $10^7$ & 1 & 0 & 90 & thick   \\
M2b & $r_{h}$ & 100 & 0  & \bf{88} & 0 & $10^7$ & 1 & 0 & 90 & thick    \\
M2c & $r_{h}$ & 100 & 0  & \bf{87} & 0 & $10^7$ & 1 & 0 & 90 & thick   \\
M2d & $r_{h}$ & 100 & 0  & \bf{86} & 0 & $10^7$ & 1 & 0 & 90 & thick    \\
M2e & $r_{h}$ & 100 & 0  & \bf{85} & 0 & $10^7$ & 1 & 0 & 90 & thick   \\
M2f & $r_{h}$ & 100 & 0  & \bf{80} & 0 & $10^7$ & 1 & 0 & 90 & thick    \\
\hline
  \multicolumn{11}{c}{Binary separation dependence}\\
  \hline
M3a & $r_{h}$ & \bf{200} & 0  & 90 & 0 & $10^7$ & 1 & 0 & 90 & thick    \\
M3b & $r_{h}$ & \bf{300} & 0  & 90 & 0 & $10^7$ & 1 & 0 & 90 & thick    \\
M3c & $r_{h}$ & \bf{400} & 0  & 90 & 0 & $10^7$ & 1 & 0 & 90 & thick    \\
M3d & $r_{h}$ & \bf{500} & 0  & 90 & 0 & $10^7$ & 1 & 0 & 90 & thick    \\
M3e & $r_{h}$ & \bf{1000} & 0  & 90 & 0 & $10^7$ & 1 & 0 & 90 & thick   \\
\hline
  \multicolumn{11}{c}{Eccentricity dependence}\\
  \hline
M4a & $r_{h}$ & 100 & \bf{0.3}  & 90 & 0 & $10^7$ & 1 & 0 & 90 & thick   \\
M4b & $r_{h}$ & 100 & \bf{0.6}  & 90 & 0 & $10^7$ & 1 & 0 & 90 & thick    \\
M4c & $r_{h}$ & 100 & \bf{0.9}  & 90 & 0 & $10^7$ & 1 & 0 & 90 & thick    \\
\hline
  \multicolumn{11}{c}{Viewing (Node) angle dependence}\\
\hline
M5a & $r_{h}$ & 100 & \bf{0.9}  & 90 & \bf{30} & $10^7$ & 1 & 0 & 90 &thick    \\
M5b & $r_{h}$ & 100 & \bf{0.9}  & 90 & \bf{60} & $10^7$ & 1 & 0 & 90 & thick    \\
M5c & $r_{h}$ & 100 & \bf{0.9}  & 90 & \bf{90} & $10^7$ & 1 & 0 & 90 &thick    \\

\hline
  \multicolumn{11}{c}{Mass ratio dependence}\\
\hline
M6a & $r_{h}$ & 100 & 0  & 90 & 0 & $10^7$ & \bf{0.1} & 0 & 90 & thick    \\
M6b & $r_{h}$ & 100 & 0  & 90 & 0 & $10^7$ & \bf{0.3} & 0 & 90 & thick    \\
\hline
  \multicolumn{11}{c}{Spin magnitude dependence}\\
\hline
M7a & $r_{h}$ & 100 & 0  & 90 & 0 & $10^7$ & 1.0 & \bf{0.5} & 90  & thick    \\
M7b & $r_{h}$ & 100 & 0  & 90 & 0 & $10^7$ & 1.0 & \bf{0.95} & 90 & thick    \\
\hline
  \multicolumn{11}{c}{Spin inclination dependence}\\
\hline
M8a & $r_{h}$ & 100 & 0  & 90 & 0 & $10^7$ & 1.0 & 0 & \bf{0}  & thick    \\
M8b & $r_{h}$ & 100 & 0  & 90 & 0 & $10^7$ & 1.0 & 0 & \bf{45} & thick    \\
\hline
  \multicolumn{11}{c}{Total mass dependence}\\
\hline
M9a & $r_{h}$  & 100 & 0.0  & 90 & 0 & \bm{$10^5$} & 1 & 0 & 90 & thick    \\
M9b & $r_{h}$  & 100 & 0.0  & 90 & 0 & \bm{$10^9$} & 1 & 0 & 90 & thick    \\
\hline
  \multicolumn{11}{c}{Optical thickness dependence}\\
\hline
M10 & $r_{h}$  & 100 & 0.0  & 90 & 0 & $10^7$ & 1 & 0 & 90 & \bf{thin}    \\
\end{tabular}
\caption{Summary of the parameters of our fiducial model (M0, first row) and its 27 variants. The four categories indicated by four separate columns refer to parameters related to the disk, the binary's orbit, the BHs, and the opacity of the emitting region.  Different rows indicate exploring the impact of each parameter. The parameter being varied with respect to the fiducial model is shown in boldface.}
\label{tab:params}
\end{table}

\subsection{Model parameters}

In total we performed a suite of 28 simulations.
The full list of model parameters that we will vary can be grouped in three sets; namely the orbital parameters: periapsis $a_{\rm per}$, eccentricity $e$, orbital inclination $i_{\rm orbit}$, and nodal angle $\Omega$; the BH parameters: BH mass $M$, mass ratio $q$, spin $a_{\rm p/s}$, BH spin inclination $i_{\rm BH, p/s}$; and the emission parameters: opacity $\tau$, and the inner radius $R_{\rm inner}$. A cartoon of the geometry of our model configuration is shown in Fig. \ref{fig:cartoon} and the various models and their parameters are summarized in Table~\ref{tab:params}.

We compute the synthetic images at four frequencies uniformly spaced between 2.5 keV and 10 keV, we focus on the X-ray emission since it is typically produced close to the BH horizon. Light curves are then generated by integrating the images and taking the sum of the total fluxes at each frequency.
In what follows light curves show the 2.5 keV emission. The initial resolution is chosen such that the images and light curves are well resolved spatially and in time when three levels of resolution are used. We found empirically that a uniform base resolution of $N^2$ pixels is sufficient for all our models when $N$ is given by
\begin{equation}
    N = 250 \left(\frac{a_{\rm per}}{ 100 R_g}\right)  \left(\frac{1 + e}{ 1 - e}\right)  \left(\frac{1}{q}\right).
\end{equation}

Our highest-resolution model is M4c, with an effective resolution of $19,000^2$ in each direction; with "effective resolution" we mean the resolution if the image was sampled uniformly with the resolution of the highest refined block. In reality the adaptive resolution and only computing the blocks close to the BHs allows us to only compute a small fraction of the effective resolution. Light curves are sampled using 1000, 5000, or 10,000 points depending on how finely detailed we find the temporal structure to be. A sampling of 5000 points is used for the eccentric models and the models with separations of 500 and 1000 $R_{\rm g}$. The sampling of 10,000 points is used for models M9 and M10 since they have the sharpest temporal features generated by the photon ring (see discussion below). Even with the optimization from adaptive resolutions and only computing blocks close to the BH position, the models are computationally expensive due to the large volume of images that needed to be computed, namely almost 400,000 in total (77,000 points at 5 frequencies), the total computational cost of the project exceeded one million CPU hours.

\section{\label{sec:res}Results}
This section reports the results of our synthetic images and light curves for all the model parameters described in the previous section.

\subsection{Fiducial model}
\begin{figure}[ht!]
      \centering
  \includegraphics[width=0.47\textwidth]{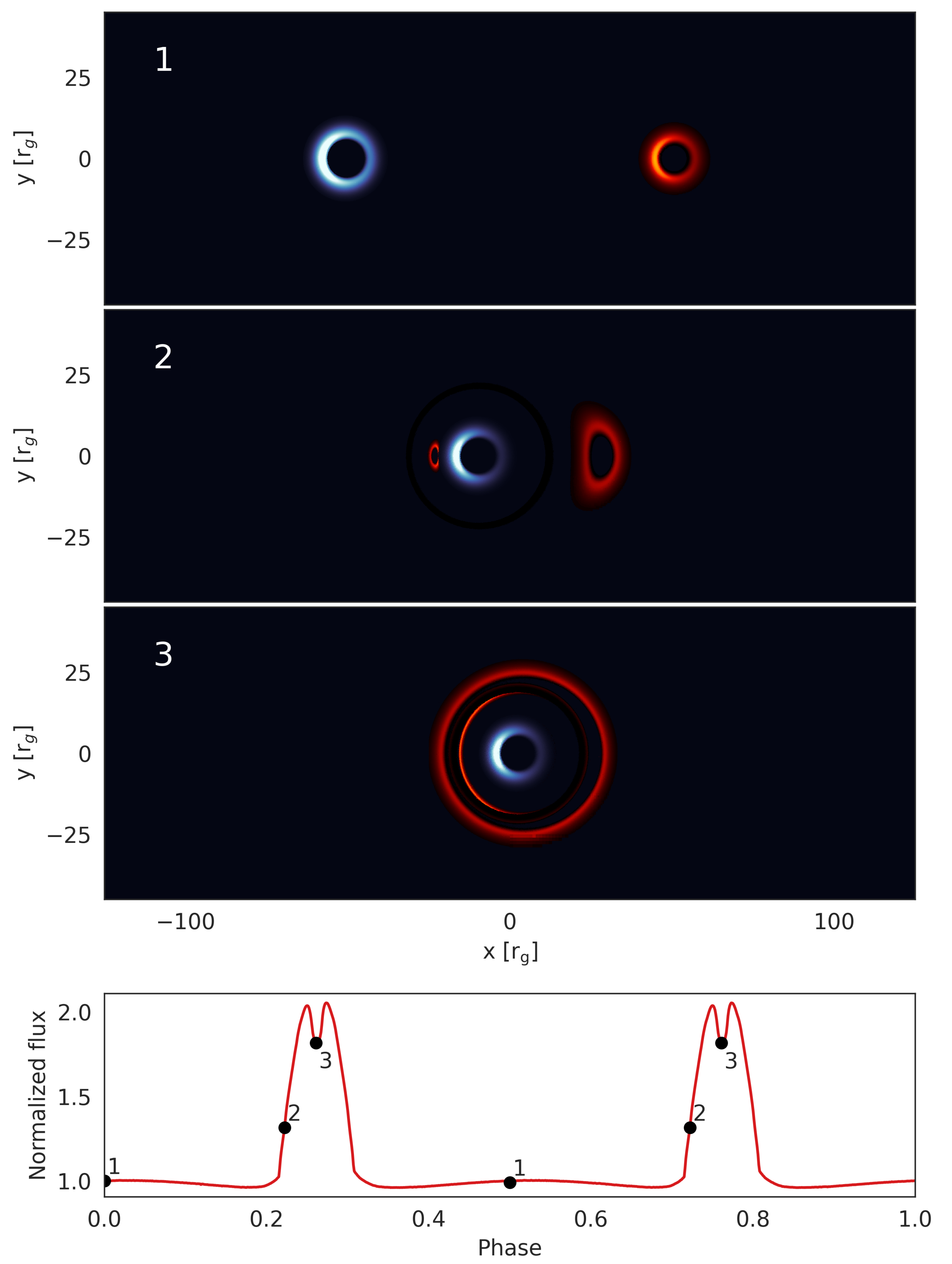}
  \caption{Light curve and images in the fiducial model. The top three panels show images of the binary when they have the largest projected separation on the sky (top panel), and on either side of the lensing event (two middle panels). Blue colors indicate an approaching BH, orange the receding BH. Bottom panel: the combined light curve of the binary, with numbers indicating the moments shown in the upper three panels.}
    \label{fig:fiducial}
\end{figure}

Snapshots of the apparent images and the full light curve in the fiducial model can be seen in Fig. \ref{fig:fiducial}. The first three panels show snapshots at three different orbital phases. The first panel shows the binary at half of the period, when the two BHs align perpendicular to the line of sight. The second and third panels show the binary at the beginning and the maximum of a SLF. The blue and red colors in the top three panels indicate emission by the minidisk of the approaching, blueshifted, and receding, redshifted BH. The bottom panel shows the total light curve at 2.5 keV, combining the emission of both BHs, where the numbers indicate the moments shown in the top panels.

At a quarter and three-quarters of the orbit, a SLF is present in the light curve. This phase corresponds to the moment when the two BHs are aligned along the line of sight. The dip visible on top of the peak of the light curve indicates that spatial variations in the minidisk emission morphology can be discerned from these lensed light curves. This is discussed in more detail in a companion paper~\cite[][hereafter Paper~II]{davelaar2021}.

To understand how the inclusion of general relativitistic effects, special relativistic Doppler boosts, and retarded times affect our light curves, we incorporate these effects one at a time into our model.

We start with a comparison between light curves generated with microlensing and from GRRT. In the literature, self-lensing is often approximated by the amplification factor $A$ derived from microlensing, the so-called "Paczynski curve"~\citep{paczynski1986},
\begin{equation}
    A = \frac{u^2 + 1}{u\sqrt{u^2+4}}.
    \label{eq:micro}
\end{equation}
Here $u\equiv b/r_{\rm E}$, with $b$ the offset between the lens and the unlensed source position on the sky, and $r_E$ is the Einstein radius. The assumptions made in deriving this simple formula is that the deflection angles are small, and that the size of the source is much smaller than the Einstein radius, so it can be treated as a point. We generate light curves by moving the lens in front of 1) a point source, 2) a Gaussian profile, and 3) a single BH image, where we use the same orbital and BH parameters as in model M0.

\begin{figure}[ht!]
  \centering
  \includegraphics[width=0.4\textwidth]{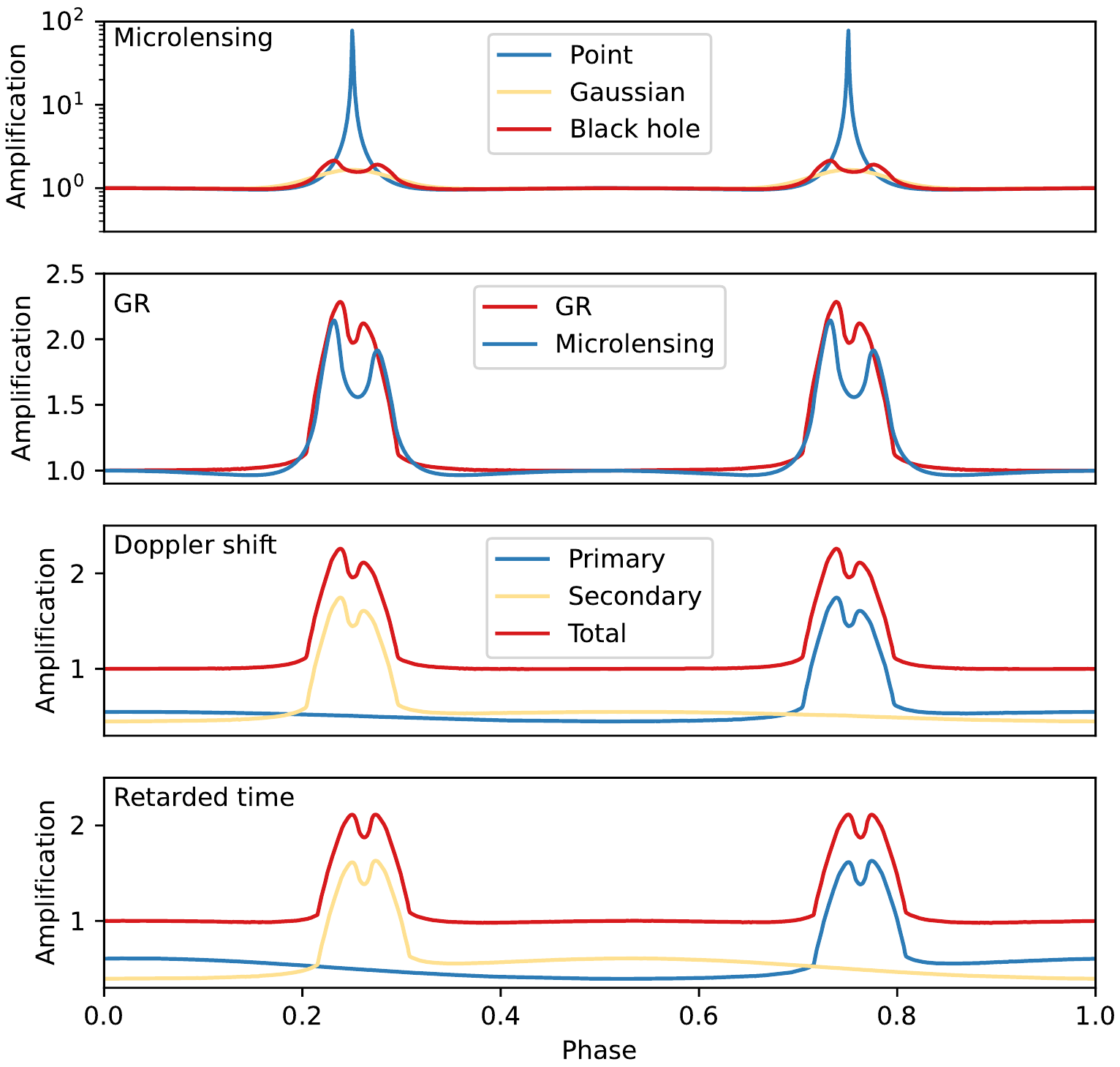}
  \caption{Various light curves showing the impact of different physical effects. The top panel shows the difference between microlensing models, assuming either point source, a source with a Gaussian surface brightness distribution, or the actual source BH image in our model. The second panel shows a comparison between microlensing and GR-generated light curves. The third panel shows the effect of relativistic Doppler boosting. The bottom panel takes all effects into account by also including time-delays.}
    \label{fig:comp}
\end{figure}

We apply the microlensing amplification to every pixel on the image plane and then compute the total flux for every point along the orbit. The resulting light curves can be seen in the top two panels of Fig. \ref{fig:comp}. The point source model shows a strong peaked flare since as the source size goes to zero equation~(\ref{eq:micro}) is singular (yielding the Einstein ring). When we increase the source size by using a Gaussian profile, the peak flattens, caused by the emission being spread out compared to the focal point of the lens, and the total amplification drops. The Gaussian has a width of $\sigma=10 R_{\rm g}=0.5 r_{\rm E}$. When we use the BH image of our fiducial model, a double-peaked flare structure appears, similar in shape to the full GRRT-generated light curve shown in the second panel. The deviations between microlensing and GRRT are caused by some light-rays suffering large deflections.

\begin{figure}[ht!]
  \centering
  \includegraphics[width=0.37\textwidth]{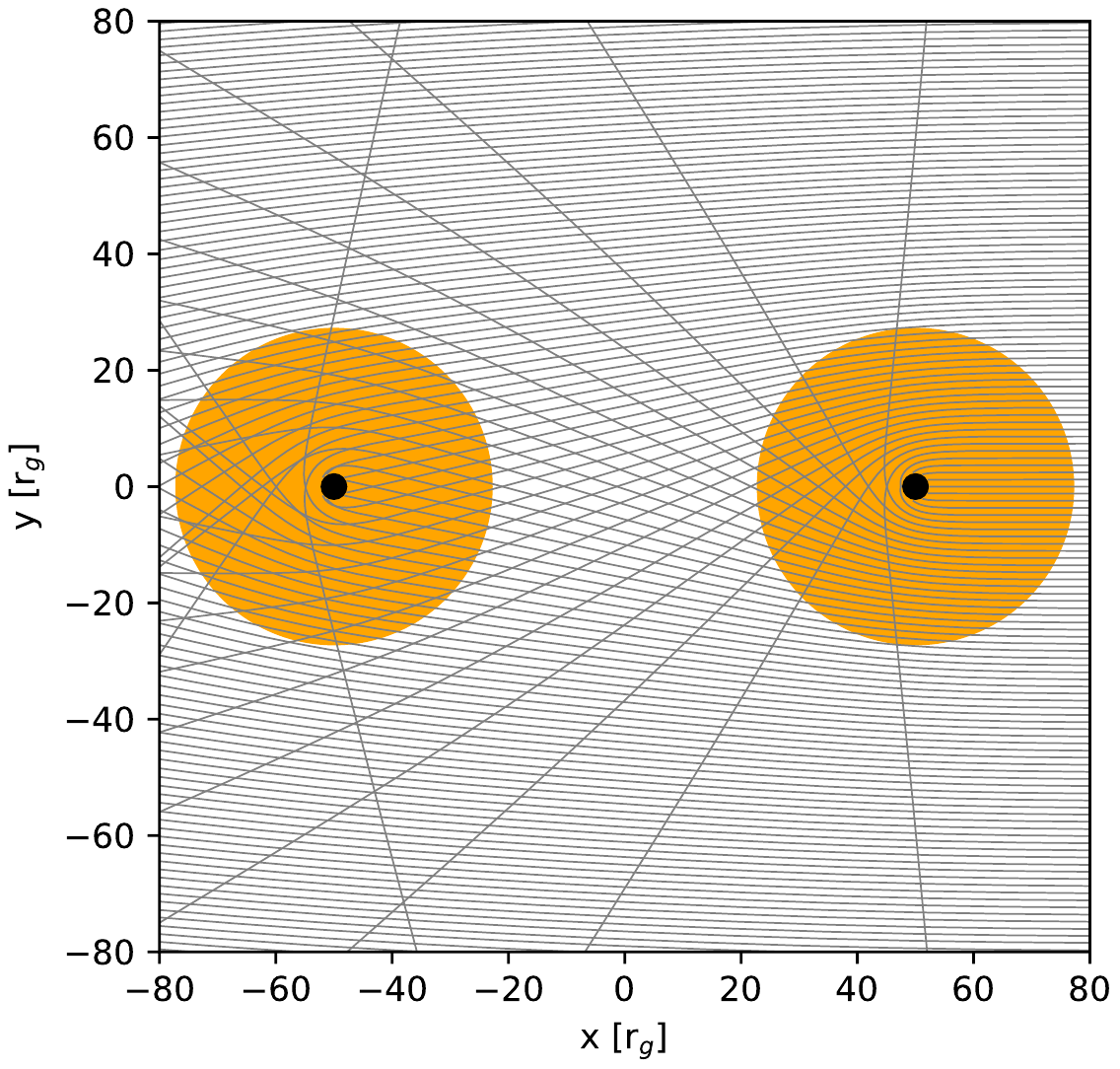}
  \caption{Geodesics in gray as an illustration of strong and weak deflections. Orange disks indicate the minidisks around each BH and black circles mark the event horizons. The observer is to the right, so that the emission from the left-side BH is lensed by the right-side BH. The rays originating from the source at positive $y$ and passing the lens at negative $y$ suffer strong deflections and form a secondary image whose brightness is not predicted accurately by microlensing.}
    \label{fig:strong}
\end{figure}

Within the Einstein ring, a secondary image of the lensed BH is visible, as can be seen in panel (2) of Fig.~\ref{fig:fiducial}. This image is associated with strong deflections. To illustrate this, in Fig.~\ref{fig:strong} we plot geodesics in the $x-y$ plane of the binary. The foreground BH, as well as the observer are on the right (positive $x$). Close to the lens (within several $R_g$), there is a subset of geodesics visible, which show large deflection angles. Geodesics originating from positive $y$ values reach the lensed BH at negative $y$ values, and result in a secondary image of the source inside the Einstein ring. Large-angle deflections are not captured by microlensing, and increase the overall amplification since a larger fraction of the source's emission reaches the observer.

Next, in addition to switching to GRRT we also add (special) relativistic Doppler shifts caused by the Keplerian velocity of the orbital motion affect the light curves. In this case, no retarded time is taken into account. Hence, at every phase, we generate an image by ray-tracing over a static spacetime geometry: as light propagates, the BHs do not move. Ref.~\cite{ingram2021} argues that for $q=1$ binaries, the line-of-sight velocity components are equal but have opposite sign, and therefore Doppler modulations cancel. This, however, is only correct if the underlying spectrum of the emission is close to a power-law; otherwise Doppler boosting and de-boosting at velocities $\pm v$ do not have the same magnitude. In the case of a power-law the boosting is given by $\frac{\Delta F}{F}=(3-\alpha) v$, where $\alpha$ is the slope of the power-law. Light curves of our fiducial model with and without Doppler shift can be seen in the third panel in Fig.~\ref{fig:comp}. We also computed the individual light curves of the BHs to show how the modulation in the total light curve is generated. We see that the overall Doppler modulation is out of phase between the two BHs and of similar amplitude resulting in almost no modulation in the overall light curve. However, some deviations can be seen, especially during the flares. Part of the receding BH disk enters the Einstein angle before it reaches zero line-of-sight velocity, therefore, there still is some Doppler de-boosting resulting in a decrease in the first sub peak of the flare. When the BH emerges from the Einstein angle it already gained some approaching line-of-sight velocity, and the resulting Doppler boosting increases the second sub-peak. Together this generates a slightly {\it less} asymmetric peak profile compared to the light curve in panel two. For unequal mass binaries Doppler or unequal accretion rates, Doppler modulations will be more clearly visible.

Lastly, we also computed our models the effects of retarded time included (with "slow light"). In this case, the BHs move as light propagates. The effect of this can be seen in the bottom panel of Fig. \ref{fig:comp}. Again, we plot the total combined light as well as the individual light curves for each BH. The individual BHs show again a modulation that we identify as Kerr-Doppler boosting \citep{Cisneros2012} (which is not limited to spinning BHs only). Which is the Doppler effect caused by the movement of the BHs, causing the approaching BH to be blueshifted, while the receding BH is redshifted. Similar to the relativistic Doppler boosting, this effect also nearly cancels in the total light curve. But similarly to the relativistic Doppler case the peak profile is less asymmetric compared to the non-boosted flares.
\subsection{Inner radius dependence}
\begin{figure}[ht!]
  \centering
  \includegraphics[width=0.4\textwidth]{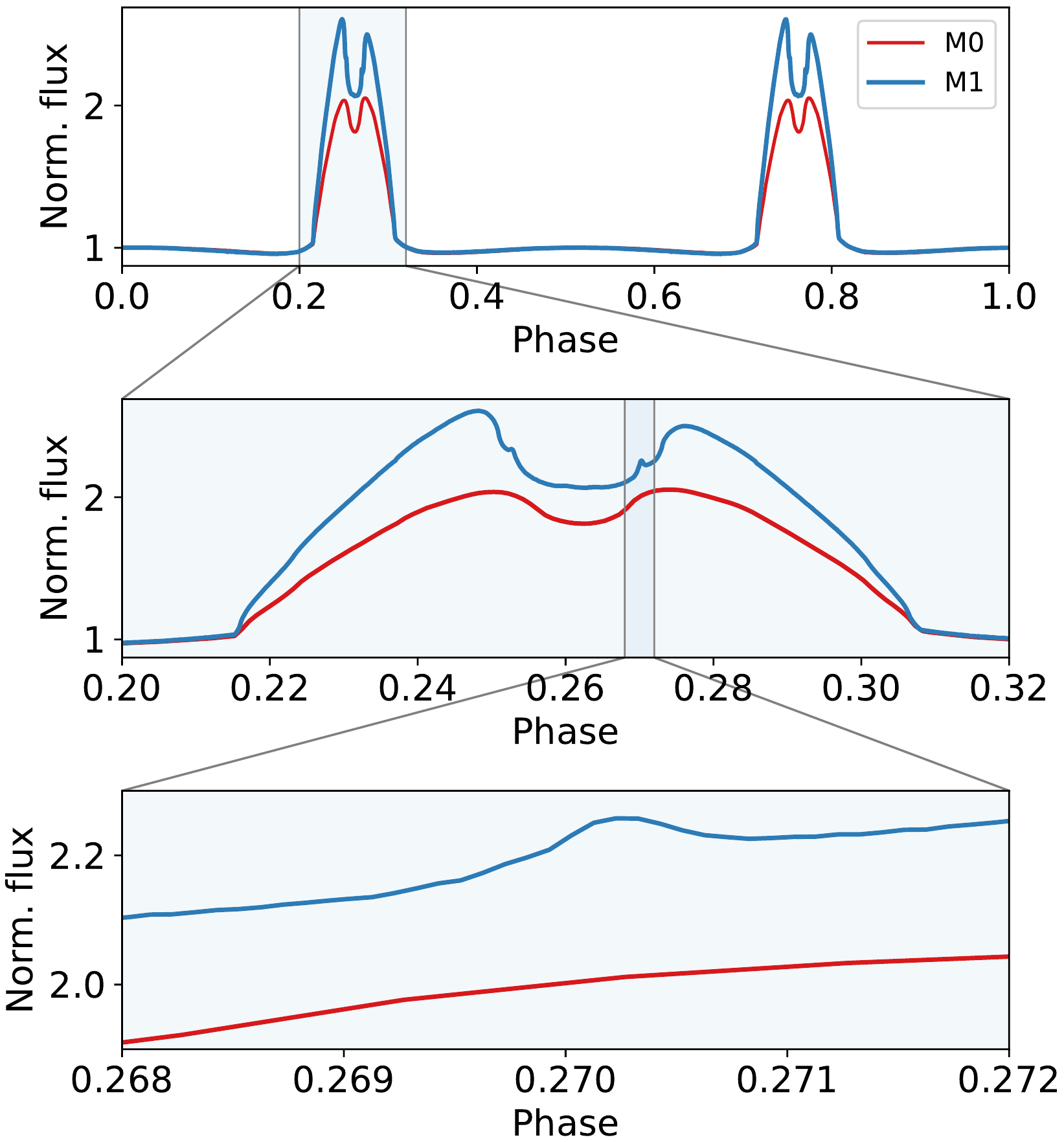}
  \caption{Light curves showing the inner-radius dependence by contrasting the fiducial model M0 (with the inner radius at the event horizon) and its variant M1 (with the inner radius moved out to ISCO). As the inner radius increases, there is a larger central gap in the accretion flow, extending outside the horizon. This enhances the amplitude of the flare, with a wider and broader "dip" (top two panels).  Additionally, the ISCO is outside the photon sphere, resulting in a photon ring in the image plane of the source. The photon ring adds an extra minor increase to the light curve on either side of the central dip (bottom panel).}
    \label{fig:nov}
\end{figure}

In our fiducial model, we chose the inner radius of each minidisk to be at the event horizon.  The motivation for this is that recent numerical works \citep{bronzwaer2021} show that in the case of geometrically thick flows, radiation inside ISCO does not necessarily vanish.
On the other hand, radiation from inside ISCO is small for geometrically thin Novikov-Thorne disks.  To understand how the inner radius alters the light curves, we computed model M1 with the inner radius at the ISCO. This model M1, as well as our fiducial model M0, can be seen in Fig.~\ref{fig:nov}.  The overall amplification in model M1 is, somewhat counter-intuitively, higher than in M0.  This is because after excising the inner region, the emission region becomes  concentrated in a narrow annular ring around ISCO, whose effective width is smaller than in the fiducial model. Also, the dip in the SLF is broader since the central gap in the accretion flow is larger compared to our fiducial model. Secondly, there are two additional features visible in the SLF, on either side of the central "dip", which are caused by the presence of the "photon ring". In the M0 model, the optically thick disk prohibits rays from circling the hole, but in the M1 model, the rays can circle ones or multiple times around the BH, an observer far away from the black hole will measure these photons appearing from a critical closed curve in the image that generates a ring like feature \citep{hilbert1917,bardeen1973,luminet1979,falcke2000b}. When the photon ring moves through the focal point of the lens, a small increase in amplification is visible. We anticipate that higher-order photon rings could also become visible if an even higher image and time resolution is used. These features are highlighted in the bottom panel of Fig.~\ref{fig:nov}, and are discussed in more detail in Paper~II.

\subsection{Binary orbital inclination dependence}
\begin{figure}[ht!]
  \centering
  \includegraphics[width=0.4\textwidth]{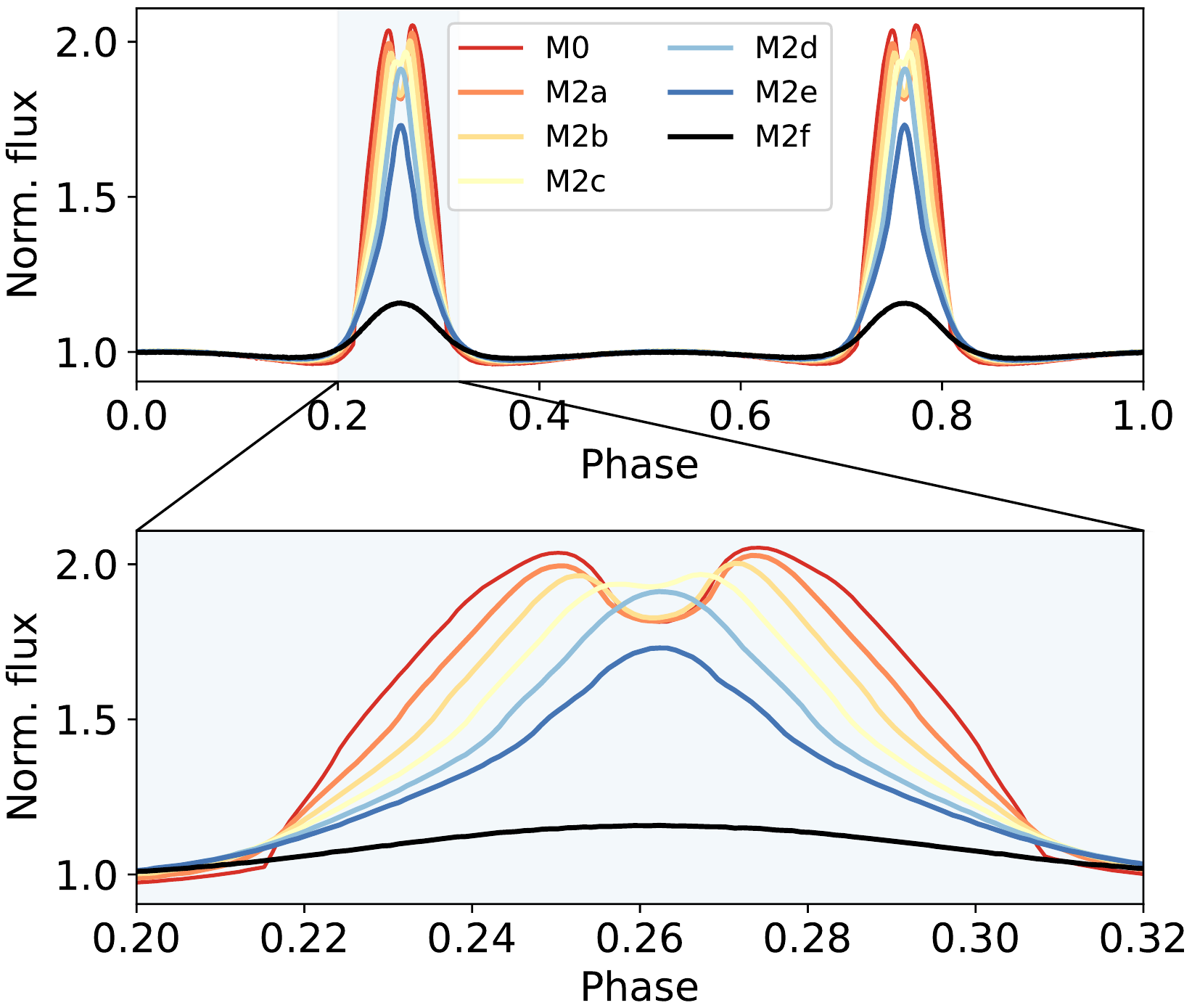}
  \caption{Light curves showing the dependence on the observer's viewing angle. The fiducial model M0 has an inclination of $90^\circ$ (edge-on), and the models M2a-f have smaller inclinations of $89^\circ$, $88^\circ$, $87^\circ$, $86^\circ$, $85^\circ$ and $80^\circ$ respectively. As the inclination decreases, the separation on the image plane between the source and the lens increase, resulting in a lower overall amplification, as expected from microlensing-based models~\cite{dorazio2018}. While the overall flare remains visible even for disks misaligned by $10^\circ$, the central dip disappears once the BH shadow's projected offset from the lens is too large (that is, in the models M2d,e,f which are more than $\sim3^\circ$ from edge on).}
    \label{fig:incl}

\end{figure}

In Fig. \ref{fig:incl} we compare models M2a-f with the fiducial model. These models differ in the inclination angle between the observer and the angular momentum vector of the binary. The inclination angles for models M1a-f are $89^\circ$, $88^\circ$, $87^\circ$, $86^\circ$, $85^\circ$ and $80^\circ$ respectively, while the fiducial model is seen edge-on at $90^\circ$. As a function of inclination, the height of the lensing flare decreases. This agrees with the expectation from microlensing, where the amplification factor depends only on the offset between the source and the lens (in units of the Einstein radius).
The offset increases with decreasing inclination, and therefore, the amplification efficiency drops, as already demonstrated by~\cite{dorazio2018}. Still, the overall flare remains visible even for disks misaligned by $10^\circ$. On the other hand, the dip in the SLF disappears for inclinations smaller than $87^\circ$, and is visible only when the focal point moves over the BH shadow. This is discussed in more detail in Paper~II.
\subsection{Separation dependence}
\begin{figure}[ht!]
  \centering
  \includegraphics[width=0.4\textwidth]{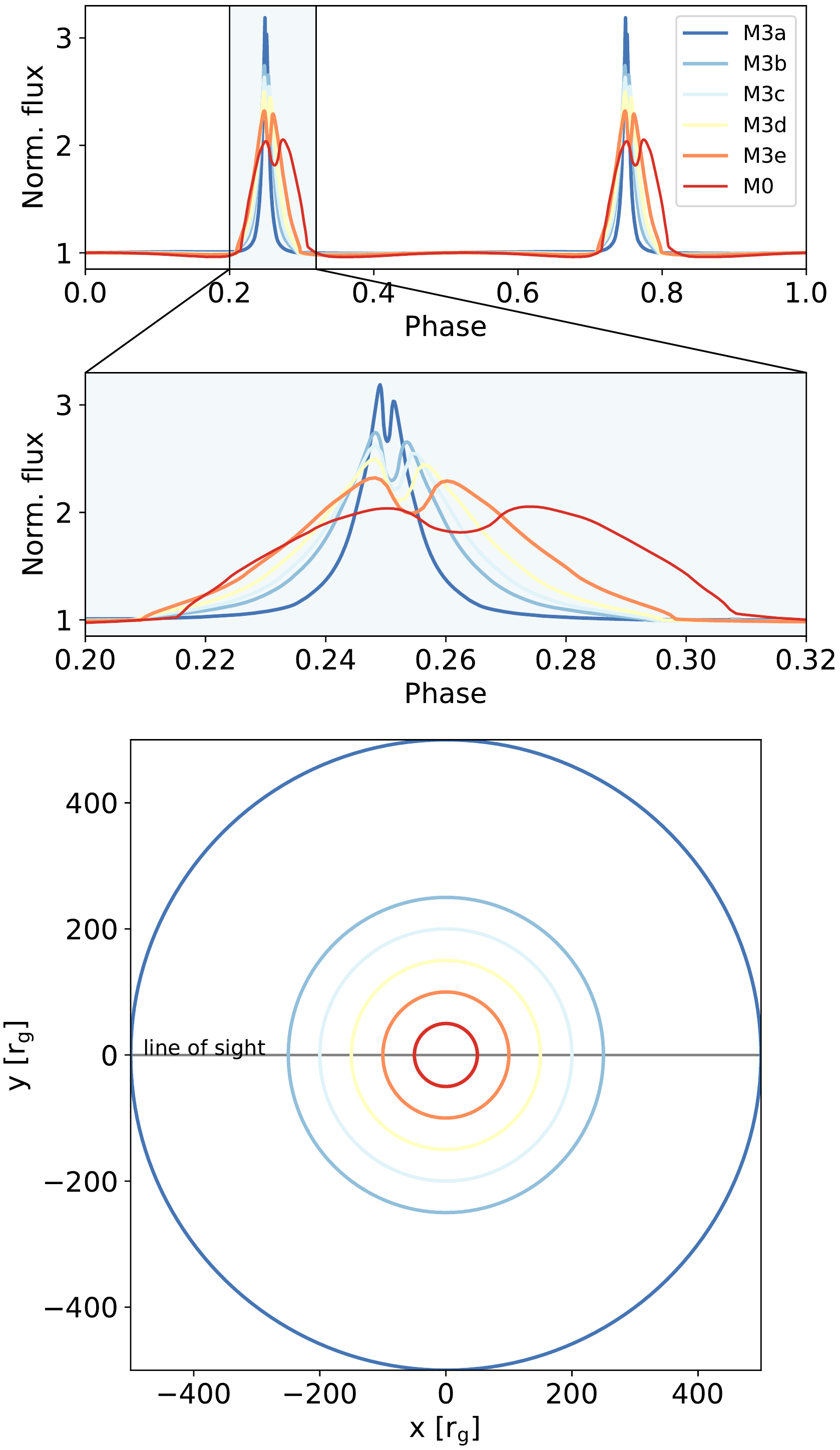}
  \caption{Light curves showing the dependence on the separation of the two BHs.  The models M3a-e have separations of $200$, $300$, $400$, $500$ and $1000~R_{\rm g}$, compared to $100~R_{\rm g}$ in the fiducial model.  As the separation increases, the angular size of the source becomes smaller, resulting in a larger amplification since a larger fraction of the source falls inside the Einstein angle, as well as a narrower width in phase, since the source spends a smaller fraction of the total orbit behind the lens, however since the period does grow, the width becomes large in physical time.}
  \label{fig:dist}
\end{figure}

In Fig.~\ref{fig:dist} we show light curves for models M3a-e, which differ from the fiducial model by varying the binary separation. While the fiducial model has a separation of $100~R_{\rm g}$, in M3a-e, this is increased to $200$, $300$, $400$, $500$ and $1000~R_{\rm g}$, respectively. In the bottom panel, we also show the orbit of each model. With increasing separation, the angular size of the source on the sky as seen by the lens decreases, and therefore, the light curve asymptotes to that of a point source model. Since the angular size of the emission region of the source shrinks, more emission falls inside the Einstein angle at the moment of maximum amplification, which results in a higher overall amplification, similar to the top panel of Fig.\ref{fig:comp}. The width of the flare also decreases with increasing separation, since the source spends less time within the Einstein angle relative to its orbital period. With increasing separation, the sizes of the individual minidisks also increase, because the tidal truncation radius becomes larger. However, this effect is minimal for an edge on view, since the source emission morphology is dominated by the strong lensing region around the BH (see the top panel of Fig.~\ref{fig:comp}).

\subsection{Eccentricity and node angle dependence}
\begin{figure}[ht!]
  \centering
  \includegraphics[width=0.4\textwidth]{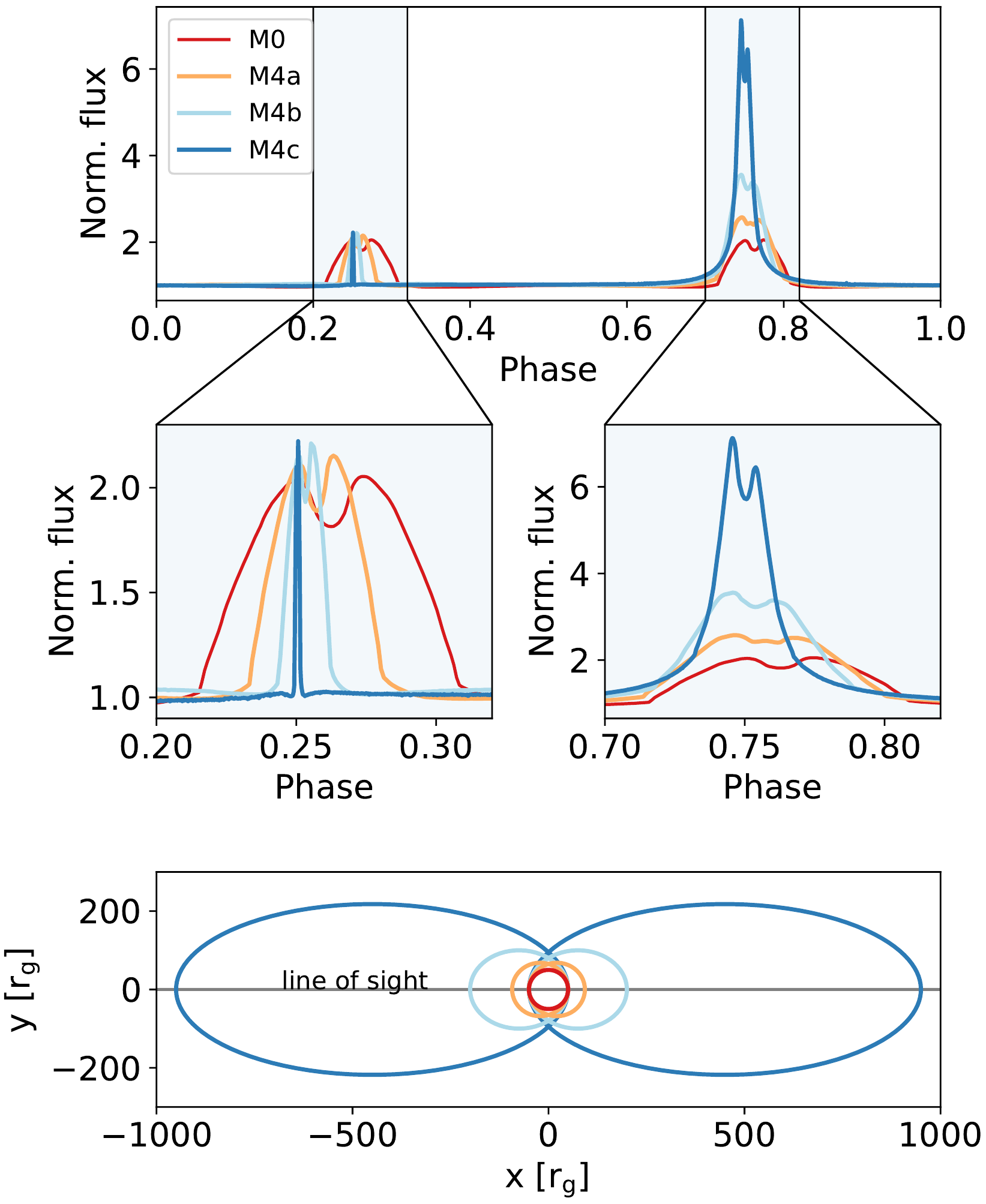}
  \caption{Light curves showing the ccentricity dependence. The fiducial model is circular, while models M4a-c have increasing eccentricities of $e=0.3, 0.6$ and 0.9, respectively.   The pericenter distance is kept constant in all four models. The bottom panel shows the orbits in the $x-y$ plane, with the observer to the right at $y=0$.The more eccentric the binary, the larger the separation at apoapsis, and the smaller the angular size of the source, this results in a higher amplification of the flux, similarly as for models M2a-e in Fig.~\ref{fig:dist}. At periapsis, the orbital velocity increases with $e$, and therefore the source spends a smaller fraction of the orbit directly behind the lens, making the flares narrower.
  }
\label{fig:ecc}
\end{figure}

In Fig. \ref{fig:ecc} we show the dependence of the light-curves on orbital eccentricity. While the fiducial model is circular, models M4a-c have increasing eccentricities of $e=0.3$, $e=0.6$ and $e=0.9$, respectively.  Note that in all models, we keep the pericenter distance the same.
In the bottom panel, we show the shape and orientation of the orbit in each model. The eccentricity of the orbit introduces a clear asymmetry between the two SLFs. One of the peaks is larger in height compared to our fiducial model. When the larger SLF occurs, the BHs are at apoapsis, while the BHs are at periapsis at the second, smaller SLF. At apoapsis, the separation is large, and the lensed BH has a smaller angular size on the sky, which results in a larger amount of emission closer to the focal point and causes the amplification to increase. This is similar to what we found for our models M3a-e. At periapse, the binaries have the same separation, so the flare amplitudes remain similar. However, the higher velocities in the more eccentric cases result in a narrower peak at periapsis than in the fiducial case, since the BH spends less time within the Einstein angle.

\begin{figure}[ht!]
\includegraphics[width=0.4\textwidth]{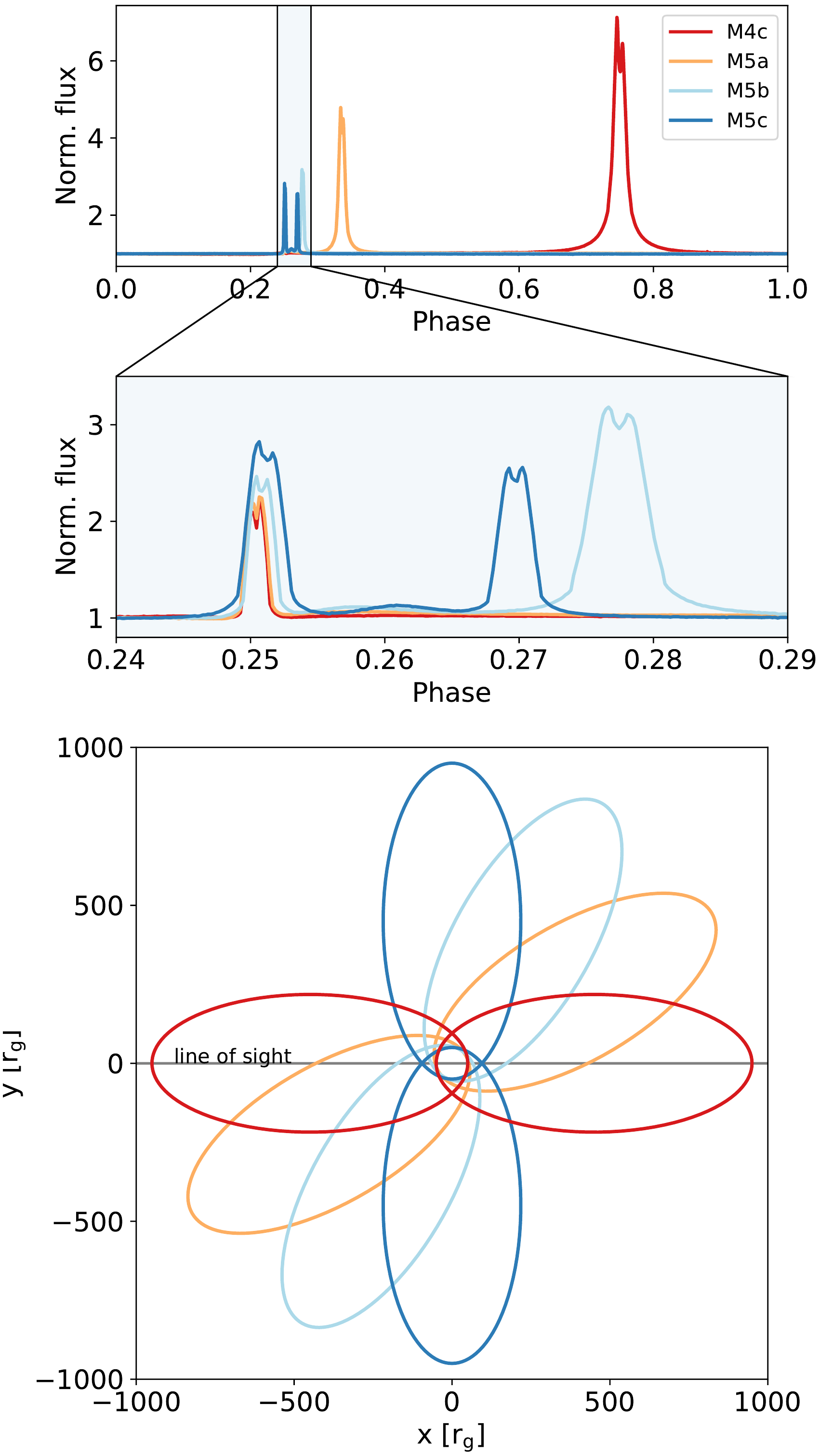}
\caption{Light curves showing the node angle dependence. All curves have eccentricities of $e=0.9$, as in model M4c, but models M5a-c change the nodal angle ($\Omega=0$ in model M4c)  to $\Omega=30, 60$ and 90$^\circ$, respectively. The bottom panel shows the orbits in the $x-y$ plane, with the observer to the right at $y=0$. As the nodal angle changes, the BHs align at different phases along the orbit and align with the minor axis in the case of model M5a. Since the lensing events happen at the closest approach here, the spacing becomes non-uniform compared to M3a. Since the separation decreases, the amplification drops, and the width narrows due to higher velocities.
  }
    \label{fig:nodal}
\end{figure}

Since the orbit is no longer axisymmetric due to the eccentricity, we also vary the nodal angle, to orient the orbital ellipse differently with respect to the line of sight. The light curves of the models M4c ($e=0.9$ and $\Omega=0$) and M5a-c (all with $e=0.9$ but $\Omega=30, 60$ and 90$^{\degree}$, respectively) can be seen in Fig.~\ref{fig:nodal}. The bottom panel again shows the orbital shape and orientation, with the observer to the right at $y=0$. As we increase the nodal angle, the height between the two SLFs becomes comparable since the difference in separation at the two flares decreases, as can be seen in the bottom panel of Fig.~\ref{fig:nodal}. The spacing between the SLFs is now also non-uniform in orbital phase or time, since the two lensing alignments occur at different phases along the binary orbit.

\subsection{Mass ratio dependence}

\begin{figure}[ht!]
  \centering
  \includegraphics[width=0.37\textwidth]{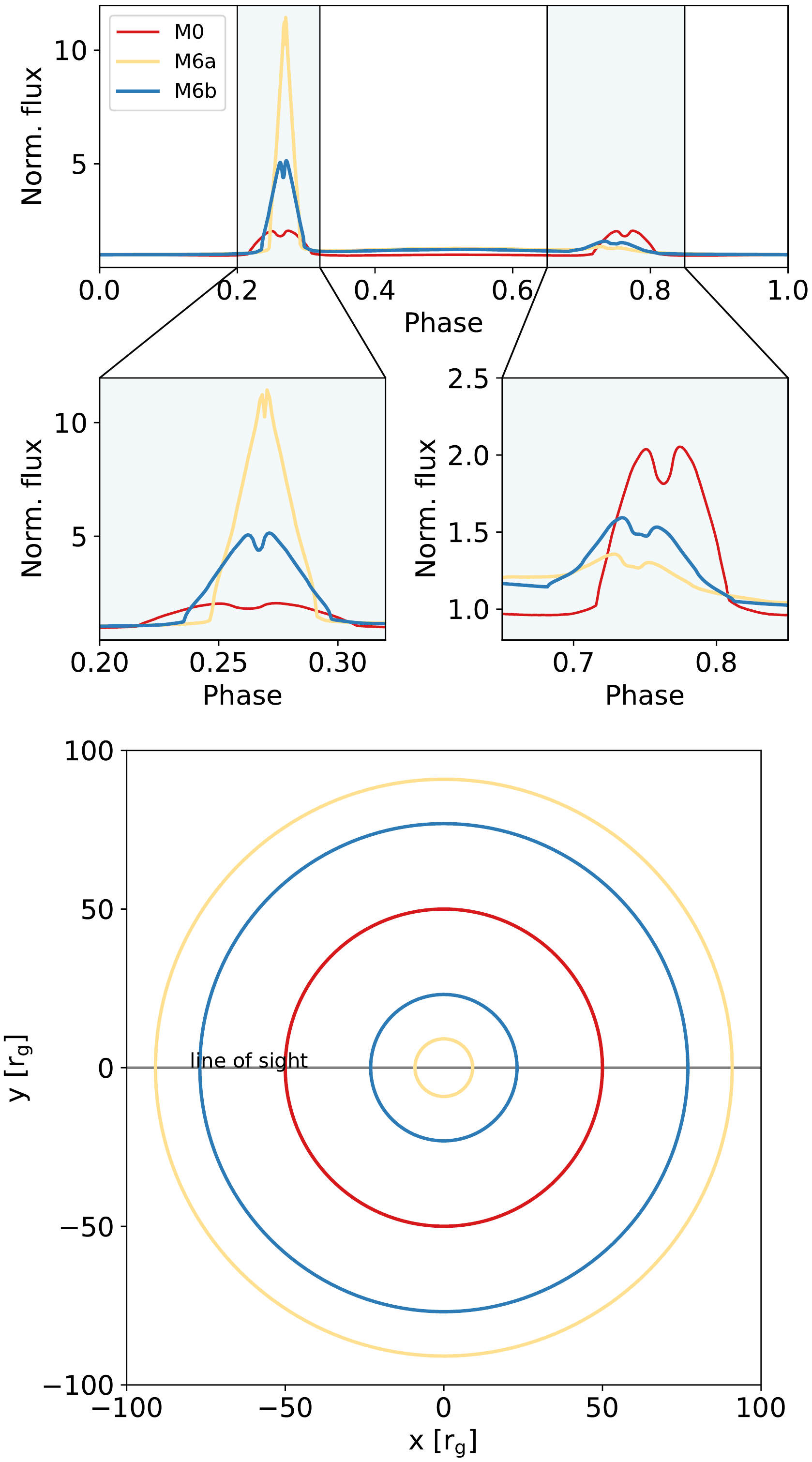}
  \caption{Light curves showing the mass-ratio dependence. Models M6a and M6b differ from the fiducial model ($q=1$) by having $q=0.1$, and $q=0.3$ respectively. As the mass ratio decreases, the angular size of the secondary BH's disk  decreases, which results in a higher amplification factor when it is being lensed. Simultaneously, the ratio between the primary source size and the Einstein radius of the secondary as the lens becomes larger, resulting in lower amplification and widening the flare profile when the primary BH is lensed.}
\label{fig:massr}
\end{figure}

The dependence on mass ratio is shown in Fig.~\ref{fig:massr}.   Models M6a and M6b differ from the fiducial model ($q=1$) by having $q=0.1$, and $q=0.3$ respectively. The binary separation and total mass are kept constant. The bottom panel again shows the orbits in the $x-y$ plane. As a function of $q$, the two SLFs differ in shape. When the secondary BH is lensed, the flare has a smaller width since the source morphology is smaller compared to the primary BH. When the primary BH is lensed, the width of the light curve is wider than the fiducial model since the source size of the primary is larger than the secondary BH. The secondary BH also dominates the measured flux since $F\propto M^{-1}$. Since there is a different amount of flux generated by each BH, their Doppler effects no longer cancel, a significant net Doppler modulation can be seen.

\subsection{Spin and tilt angle dependence}
\begin{figure}[ht!]
  \centering
  \includegraphics[width=0.37\textwidth]{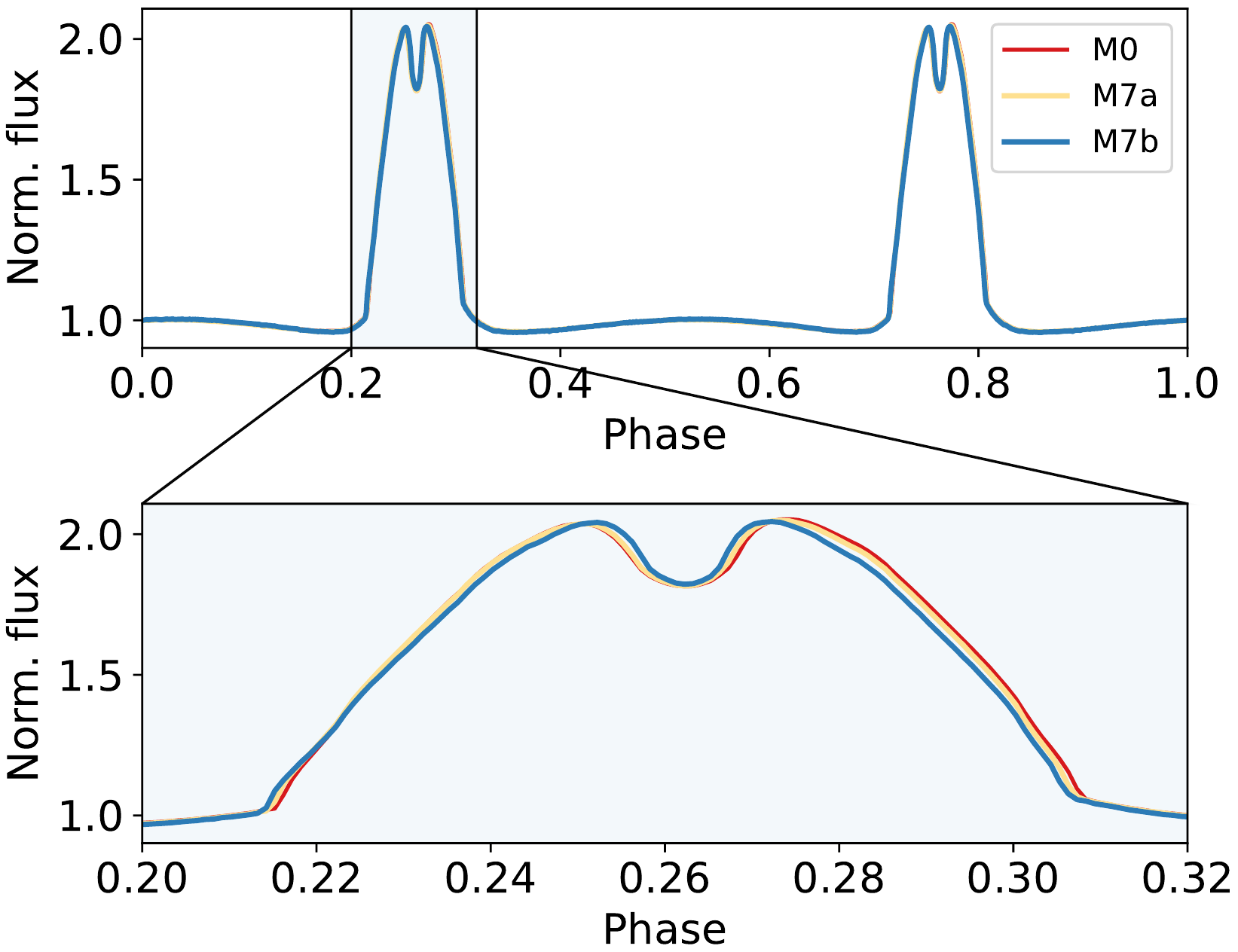}
 \caption{Light curves showing the spin magnitude dependence.  The fiducial model has zero spin ($a=0$) while models M7a-b have spins of $a=0.9$ and 0.95, respectively. The spin dependence only has a relatively small effect on the asymmetry of the double-peaked structure, since it only modifies the emission and the space-time metric within a few gravitational radii of the BHs.}
  \label{fig:spin}

\end{figure}

Fig. \ref{fig:spin} shows the BH spin-dependent models M7a-b. The fiducial model has a non-spinning BH, while for models M7a-b both BHs have a spins of $a=0.5$ and $a=0.95$, respectively. The effect of spin turns out to be small on the global properties of the light curve; this is expected since the spin does not affect the overall lensing properties to first order. Spin alters the accretion disk's inner radius and its velocity profile in the innermost regions. This results in a narrower dip in the light curve and a lower second sub-peak for each flare. Both of these effects are small, but can be discerned in the bottom panel of Fig.~\ref{fig:spin}.

\begin{figure}[ht!]
  \centering
  \includegraphics[width=0.37\textwidth]{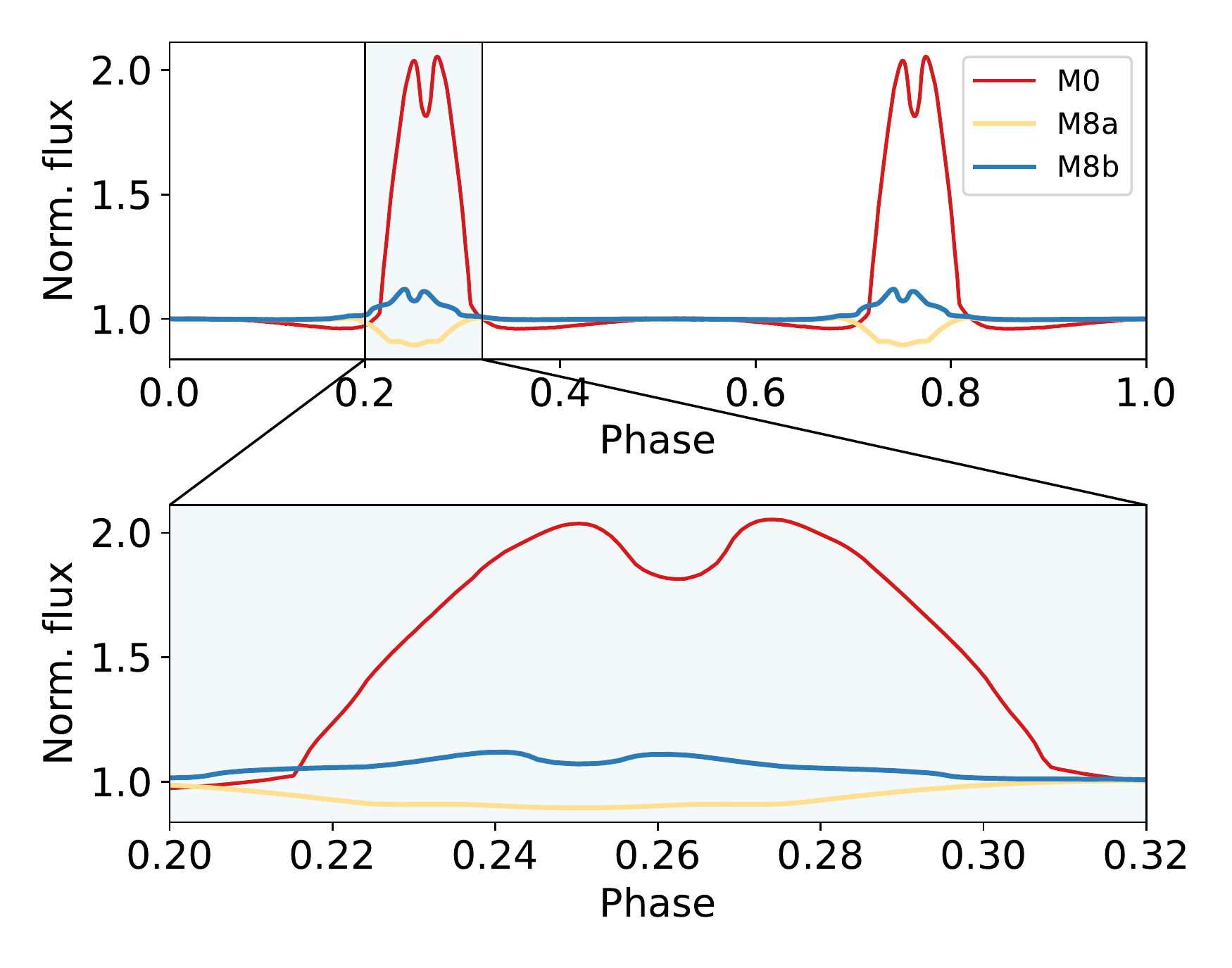}
 \caption{Light curves showing the spin orientation dependence. The BH spins and their minidisks in models M8a-b differ in their inclination with respect to the line of sight. In the fiducial model, both minidisks have $i_{\rm disk}=90^{\degree}$ (edge-on), whereas models M8a-b have $i_{\rm disk}=0^{\degree}$ (face-on) and $i_{\rm disk}=45^{\degree}$, respectively. For nearly face-on minidisks, a new effect appears, as the foreground disk can block the light from the lensed BH - this physical occultation can erase the flare or replace it with a transit-like depression.}
  \label{fig:spin-inc}
\end{figure}

We also varied the angle between the disk and the BH spin axes for models M8a-b. These models have fixed spin magnitudes of $a=0.95$ but both BHs, as well as their respective minidisks, are inclined with respect to the binary's orbital plane, with disk inclinations of $i_{\rm disk}=0^\circ$ and $i_{\rm disk}=45^\circ$, respectively. The resulting light curves are shown in Fig.~\ref{fig:spin-inc}.

When the minidisks are inclined, a new effect appears, because the foreground disk, which is assumed to be optically thick, can physically block the light from the lensed background BH. Due to this occultation, the lensing flare can be erased, or, for a face-on minidisk, replaced by a depression similar to those  see in planetary transit light-curves. This occultation was also found by \citep{ingram2021}.

\subsection{Spectral dependence}

To assess the spectral-slope dependence, we varied the mass of the BHs in models M9a-b, which have total binary masses of $10^5 M_\odot$ and $10^9 M_\odot$ respectively, compared to $10^7 M_\odot$ for the fiducial model. This shifts the peak of our spectrum either to lower or higher frequencies since $F\propto M^{-1}$. The change of the peak alters the spectral slope, the slopes are computed between 2.5 and 10 keV and are $\alpha=1.1$, $1.7$ and $-0.8$ respectively.
In the case of our fiducial and M9a model, the spectral slope is positive, while for the M9b model, the spectral slope turns to negative values. This does, however, not affect the main features we reported before. The morphology of the emission region slightly changes, causing the flare to have a lower amplitude. Since the orbit is eccentric and the BHs have an equal mass, we do not see strong Doppler modulation changes due to spectral slope difference, but we anticipate those to be more prominent in a model with a large total binary mass and unequal mass ratios, or if $i<80^{\degree}$ when the light curve is only dominated by the Doppler modulations caused by the orbit.

\begin{figure}[htp!]
  \centering
  \includegraphics[width=0.43\textwidth]{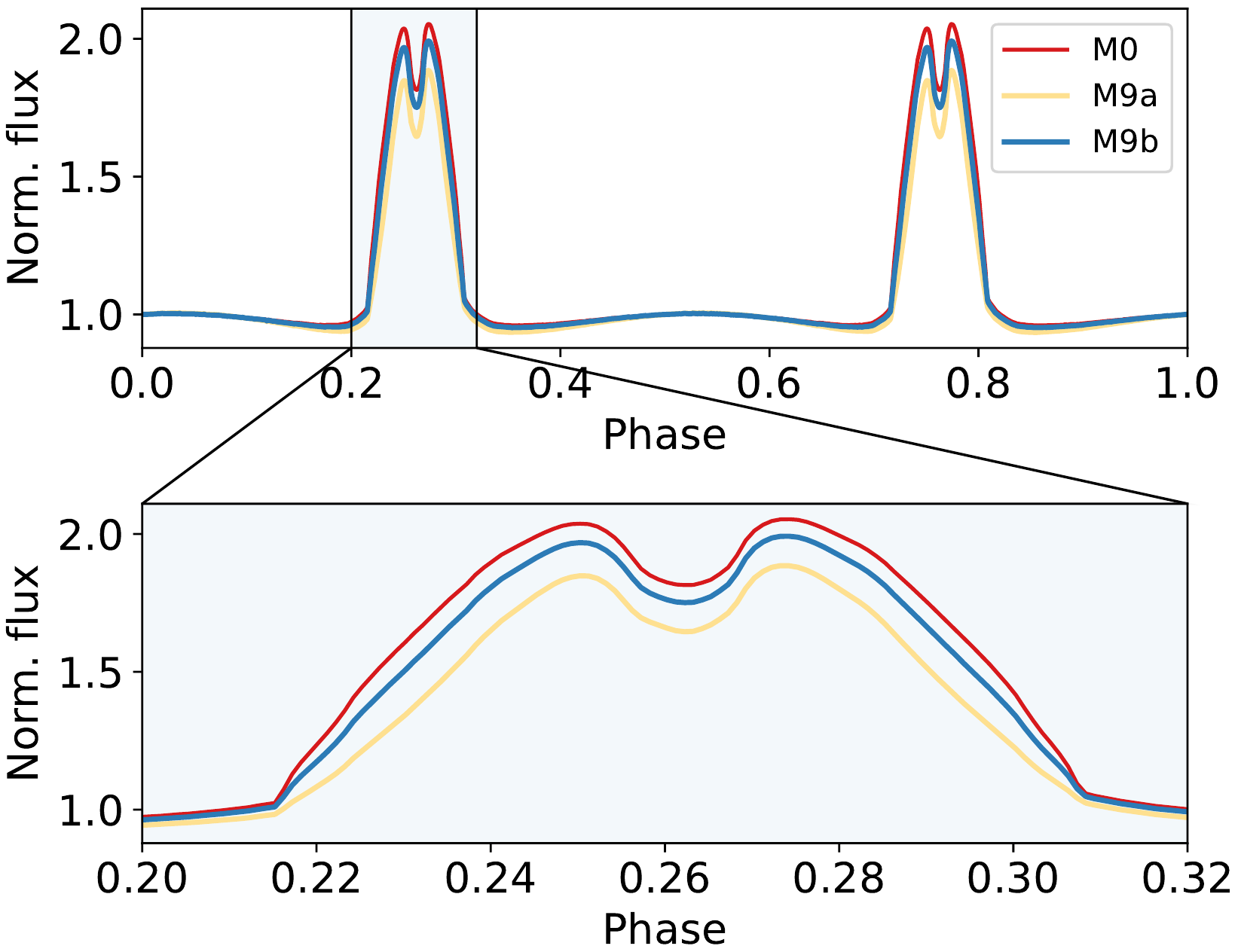}
  \caption{Light curves showing the dependence on the spectral shape, via changing the BH masses.  Models M9a-b have a BH mass of $10^5~M_\odot$ and $10^9~M_\odot$, respectively, compared to $10^7~M_\odot$ in the fiducial model.
  This changes the spectral slope in the observed X-ray band (between 2.5 an 10 keV) to $\alpha=1.1$, $1.7$ and $-0.8$ respectively.
  The spectral slope only modestly affects the overall flare amplitude. We expect it to have a larger impact on the overall Doppler modulation during the orbit in the case of an unequal-mass BH, where the Doppler effects from the two BHs do not nearly cancel.}
  \label{fig:mass-xray}
\end{figure}
\subsection{Opacity dependence}
\begin{figure}[ht!]
  \centering
  \includegraphics[width=0.43\textwidth]{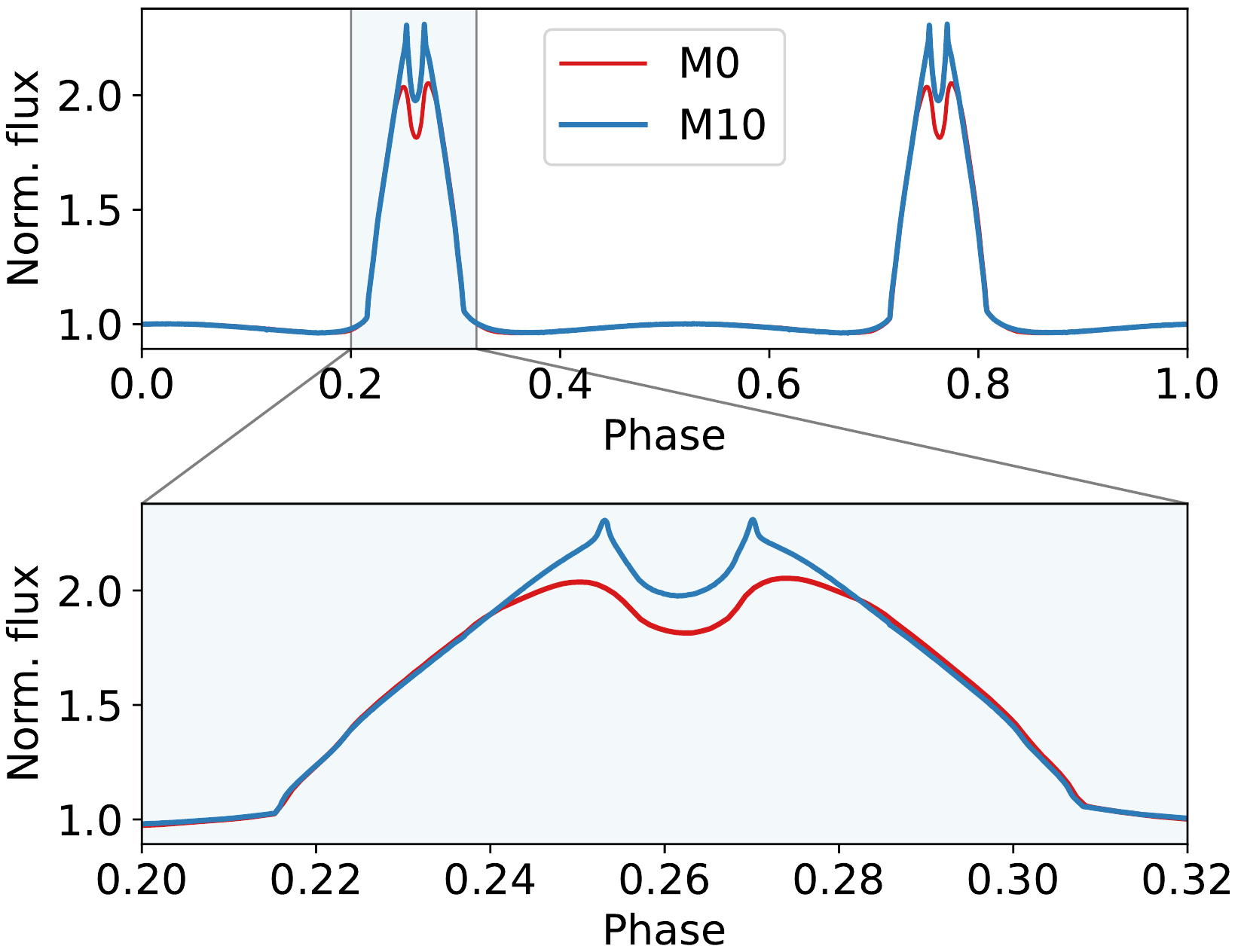}
  \caption{Light curves demonstrating the impact of optical depth. Model M10 is identical to the fiducial model M0, except the minidisks are assumed artificially to be optically thin, while in the fiducial model M0 they are optically thick.  The sharp features seen for M10 arise from the photon rings. These can only form in the optically thin case, and concentrate the emission to a small solid angle, allowing stronger amplification when parts of the bright ring are behind the lens.}
  \label{fig:thin}
\end{figure}

Finally, in Fig. \ref{fig:thin}, we show model M10, which has an optically thin disk. In order to isolate the effect of the opacity alone, we retain the same geometrically thin shape of the emission as in the other models (even if this is unphysical). Since the opacity is zero, light rays can orbit several times around the BH. This allows the photon ring to become visible, in contrast to the optically thick models. In model M10, the amplification is larger than in the fiducial model since the photon ring is a sharp feature, compressing most of the emission in the domain into a small solid angle.
%

\section{\label{sec:dis}Discussion}

In this section, we compare our results to previous works, list observational challenges for observing SLFs, and discuss possible implications of a SLF measurement.

\subsection{Comparison to previous works}
Previous work by \cite{hu2020} showed that point source models of self lensing binaries could explain the observed flare for the source {KIC-11606854}, which they dubbed as Spikey. Compared to their work, we included GR effects and finite source sizes. We find that the source size alters the width of the self lensing flare; the flare width, therefore, gives information on the emission morphology of the lensed source. Additionally, Kerr-Doppler boosting is shown to increase the overall Doppler modulation of individual BH light curves, leading to inaccurate model parameter predictions if only relativistic Doppler boosting is included. In future work, we aim to fit our presented model to Spikey's light curve in different bands.

Ref.~\citep{dorazio2018} computed self-lensing light curves by using the microlensing approximation. In contrast to \citep{hu2020}, ref.~\citep{dorazio2018} investigated the dependence on source size morphology and found that it is possible to extract information on the accretion disk structure, namely disk sizes. This is in agreement with our findings. Additionally to the work by \citep{dorazio2018}, we also include strong lensing in the vicinity of the individual BHs. This changes the source morphology since the BH lenses the accretion flow around itself. An infinitesimally thin accretion disk viewed edge-on, as considered in the work by \citep{dorazio2018}, would only show a thin line, while in our case, we see a circle with a hole in it (see top panel in Fig.~\ref{fig:AMR}). Ref.~\citep{dorazio2018} also computes light curves in the case where the minidisk is misaligned with the orbit, which  generates source morphologies that are closer to ours. However, since only very large binary periods are considered ($\sim 4$ years, with a mass of $10^6 ~M_\odot$), the angular size of the central hole in the minidisk emission is negligibly small. The dip in the SLF that we report here and in Paper~II is therefore not visible.

Compared to the work by \cite{ingram2021} we use a superposed binary metric, including dynamical evolution of the spacetime during light propagation. This results in the inclusion of Kerr-Doppler boosting. The light curves also differ in one distinct way between our work and the work by \cite{pihajoki2018}. Due to different spectral shapes, the light curves of \cite{ingram2021} are more single-peaked, and only a less prominent, minor asymmetry can be seen in their predicted light-curves. In their model, the spectrum's peak is shifted towards the UV, meaning that at harder X-ray wavebands, the spectral slope is much steeper, resulting in much stronger Doppler (de)boosting, which results in a less symmetric image morphology. Similarly to us, they assume an optically thick black body spectrum. We show that for the optically thin case, the peak width is much narrower. Both our models have in common that the underlying emission model is very simplified compared to the realistic case of blazars and other AGN. In reality, sources have multiple components contributing to the spectra, e.g., the disk, jet, and X-ray corona. Therefore, investigating how the light curves behave with more realistic spectral models is needed to fully understand the flare properties expected for these sources.

\subsection{Observational limitations}
In our work, we neglect the dynamics of the accretion flow. The viscous time scales are larger than the self lensing time scale, but it will add temporal variations on the overall light curve. Dynamical accretion models have been reported in the literature either in hydrodynamics \citep{macfadyen2008,dorazio2016} and more recently also in GRMHD \citep{dascoli2018,combi2021,lopez2021}. Future ray-tracing of GRMHD simulations of MBHBs in higher Eddington states should give more insight into the effect of accretion flow dynamics on MBHB light curves. Work by \citep{dascoli2018} shows ray traced images of a close separation MBHB and compute bolometric luminosity as a function of the azimuthal positon of the observer. For inclinations close to the edge on a flare is present that also contains hints of a dip (see their Fig. 11).

Accretion variability of AGN can also induce flares. However, these flares are not periodic and can therefore be averaged out by using phase folding. Additionally, since self-lensing is purely a geometrical effect due to light bending, it is achromatic. Therefore, amplification by self-lensing should depend on the observed frequency relatively weakly and strongly correlate between bands. In contrast, accretion-induced variations are strongly frequency dependent, and empirically, the optical and UV fluctuations are not fully correlated~\citep{xin+2020}. GRMHD also provides a  first-principle velocity profile for the accretion, which might affect the observability of the double-spiked SLF profile we report here and in Paper~II, however Fig 11 in \citep{dascoli2018} shows azimuthal profiles of the bolometric luminosity generated by ray tracing GRMHD simulation which do show a hint of a dip within the SLF for edge on configurations. In certain configurations, velocity uncertainties may be less important, e.g. if the emission is produced by an optically thin corona with small internal bulk velocity, or if the spin axis of the lensed BH is misaligned with respect to the orbital axis (so its emitting material is face-on and has small line-of-sight velocities).

To scale our results to observables, the periods of our light curves can be scaled by black hole mass. We showed that there is some mass dependence on the light curve, but this is minimal. For our fiducial model with separation of $100 R_{\rm g}$ and a total black hole mass of $M=10^7 M_\odot$, the period is roughly a day.

At fixed binary separation in gravitational units, the orbital period scales linearly with total BH mass.  As a result, for BHs up to $10^{9-10} M_\odot$, the periods reach 100-1000 days (and larger if the binary separation is also increased). In the present sample of binary quasar candidates, identified based on their optical periodicity in large time-domain surveys~\citep{graham2015,charisi2016}, the BH masses are skewed towards these high masses, and the orbital periods are comparable to a year. These candidates would be favorable for searching for SLFs. In the case of our fiducial model with a separation of $a=100~R_{\rm g}$ the period of the orbit for a $10^9 ~M_\odot$ would be approximately $0.7$ years. The SLF duration is roughly ten percent of the total orbit, so the duration of the SLF is $\sim 25$ days. In the case of model M3e, which has the largest separation among our circular models ($1000~R_{\rm g}$), we find a total orbital period of 22 years, and an SLF duration of approximately 160 days.

In our fiducial model, the binary inspiral time due to GW emission is approximately $1.5$ years (\citep{peters1964}; see Eq.~28 in \citep{haiman+2009}). This makes it hard to catch such a short-lived binary, with our fiducial parameters, in current surveys.
However, for large future time-domain surveys, such as by the Vera Rubin Observatory's Legacy Survey of Space and Time (LSST), with a few-day cadence, these rare and short-lived binaries should be present, and could be identified as ultra-short periodic sources~\cite{kelley2021,xinhaiman2021}.
The GW inspiral time scales linearly with mass (at fixed binary separation in gravitational units), increasing to as high as $\sim 10^{2-3}$ years for the binary candidates with masses of $\sim 10^9~{\rm M_\odot}$ and periods of order a year~\cite{graham2015,charisi2016}.
Since the inspiral time scales more steeply with binary separation ($T\propto a^4$), increasing the separation will improve the observability; the binaries will be longer lived, and the SLFs will be stronger, due to smaller angular source sizes. For example our model M3e with a seperation of $a=1000 R_{\rm g}$ the GW inspiral time is $15,000$ years with a binary period of $2.6$ months in the case of a $10^7 ~M_\odot$ binary and $10^{6-7}$ years with a binary period of 22 years for binary candidates with masses of $\sim 10^9~{\rm M_\odot}$.

Apart from identifying MBHBs to begin with, a clear practical obstacle to identifying lensing flares, and characterizing their shapes, including the "dips", is stochastic AGN variability. The typical X-ray variability of AGNs has an RMS value of about 10\% and 40\% over a day to week timescales \citep{lawrence1993,maughan2019}. The SLF exceeds this RMS value for inclinations larger than $80^{\degree}$. However, the dip in the SLF for edge-on viewing angles is of the order of 20\% of the total SLF amplification, although the time scales are much shorter compared to the typical accretion-induced variability. As stated before, phase folding can mitigate this problem, but a first hint of the SLF should at least stand out of accretion induced variability or system noise when only one period is observed.

\subsection{Implications}

In this work we investigated the MBHB parameter dependence on SLFs, and we report non-degenerate features that when observed will help improve parameter predictions. Observing equal-height SLFs is a strong indication of a close to equal mass binary on a circular orbit. Any deviations from this that result in the larger SLF to be narrower than the smaller SLF indicates an unequal mass binary. Eccentricity induces unequal height SLFs as well, but in this case the smaller SLF is also narrower. The nodal angle for eccentric binaries changes the relative spacing of the SLFs within the full 2$\pi$ orbital phase; this can be used to constrain eccentricity and nodal angle.  Inclination leaves is imprint by keeping the width constant but the height smaller. Separation distance sets the overall period of the orbit but also the amplification factor of the SLF. We do not find a strong dependency on spectral shape or black hole spin. If the SLF is observed close the edge-on, there is a dip visible in the SLF, caused by substructure in the source morphology and is directly related to the size of the black hole shadow, as discussed in more detail in Paper~II.

Our results also have implications for future LISA observations. If self-lensing flares are detected in the electromagnetic observations for a LISA source prior to its merger~\citep{kocsis+2008}, with the phase of the flares tracking that of GWs~\citep{haiman2017}, it will help secure unambiguous identification of the source. Current projections makes this possible for at least a fraction of LISA binaries, i.e. binaries with accurate sky positions derived from the GW inspiral signal days before merger~\citep{mangiagli+2020}.  Having the electromagnetic source identified will provide its precise sky location, and help better constrain binary parameters such as the inclination angle and distance. Secondly, the self-lensing flares always occur at the same orbital phase, and are tied directly to the GW phase through the binary's orbital motion. Therefore, they would provide a clean experiment to correlate the arrival times of GWs  and photons, which can be used to constrain graviton masses and alternative theories of gravity in which the propagation speeds of photons and gravitons differ~\citep{haiman2017}.

\section{\label{sec:concl}Conclusions}

In this study, we presented a self-lensing binary model, extending on recent work. To this end, we utilized the existing general relativistic ray tracing code RAPTOR, which we optimized by implementing an adaptive mesh refinement scheme for the camera plane. We constructed a superposed Cartesian binary metric in which the BHs are on Keplerian orbits. The emission is assumed to be produced by two minidisks surrounding each BH. We generated synthetic light curves for a variety of orbital, BH, and emission model parameters. We showed how finite source sizes alter the shape of the light curves compared to previous works that used point source modeling and recovered point source-like behavior only when the separation is sufficiently large. We highlighted the importance of GR ray tracing for modeling these systems, by finding discrepancies between microlensing and GR lensing. For gravitational waves sources, as will be measured by LISA, observing self lensing flares would help orbital constraint parameters such as the inclination and help constrain the graviton mass. Observing SLFs provides an exciting opportunity to not only constrain MBHB parameters relevant to LISA GW observations but also opens a new way to measure black hole shadow sizes in systems that are unresolvable by current VLBI facilities.

\begin{acknowledgments}
The Authors thank Christiaan Brinkerink, Anastasia Gvozdenko, Jeremy Schnittman, and Daniel D'Orazio for valuable comments and discussions during this project. JD is supported by a Joint Columbia/Flatiron Postdoctoral Fellowship. Research at the Flatiron Institute is supported by the Simons Foundation.  We acknowledge support by NASA grant NNX17AL82G and NSF grants 1715661 and AST-2006176. Computations were performed on the {\tt Popeye} computing cluster at SDSC maintained by Flatiron Institute's SCC. This research has made use of NASA's Astrophysics Data System. {\it Software:} {\tt python} \citep{oliphant2007,millman2011}, {\tt scipy} \citep{jones2001}, {\tt numpy} \citep{vanderwalt2011}, {\tt matplotlib} \citep{hunter2007}, {\tt RAPTOR} \citep{bronzwaer2018,bronzwaer2020}.

\end{acknowledgments}

%
\end{document}